\documentclass{article}
\usepackage[utf8]{inputenc}
\usepackage{textcomp}

\usepackage{amsthm}
\usepackage{amsmath}
\DeclareMathOperator*{\argmin}{arg\,min}
\DeclareMathOperator*{\argmax}{arg\,max}
\usepackage{amssymb}
\usepackage{courier}
\usepackage{enumerate}
\usepackage{algorithm}
\usepackage{algpseudocode}

\usepackage[title]{appendix}
\usepackage{float}
\usepackage{graphicx}
\usepackage{comment}
\usepackage{bm}
\usepackage[authoryear]{natbib}
\makeatletter
\theoremstyle{remark}
\newtheorem{remark}{Remark}

\newtheorem*{thm1}{Theorem 1}
\newtheorem*{thm2}{Theorem 2}
\newtheorem*{thmA.1}{Theorem A.1}
\newtheorem*{thmA.2}{Theorem A.2}

\newcommand{\keywords}[1]{\noindent\textbf{Keywords:} #1}
\newtheorem{example}{Example}
\usepackage[a4paper,left=2.5cm,right=2.5cm,top=2.5cm,bottom=2.5cm]{geometry}
\usepackage[title]{appendix}
\usepackage{authblk}

\title{Measuring Variable Importance via Accumulated Local Effects}
\author{Jingyu Zhu and Daniel W. Apley\thanks{Address for correspondence: Daniel W. Apley, Department of Industrial Engineering \& Management Sciences, Northwestern University, Evanston, IL 60208, USA. E-mail: apley@northwestern.edu}}
\affil{Northwestern University, USA}
\date{}
\begin{document}

\maketitle

\begin{abstract}
A shortcoming of black-box supervised learning models is their lack of interpretability or transparency. To facilitate interpretation, post-hoc global variable importance measures (VIMs) are widely used to assign to each predictor/input variable a numerical score that represents the extent to which that predictor impacts the fitted model’s response predictions across the training data. It is well known that the most common existing VIMs – namely marginal Shapley and marginal permutation-based methods – can produce unreliable results if the predictors are highly correlated, because they require extrapolation of the response at predictor values that fall far outside the training data. Conditional versions of Shapley and permutation VIMs avoid or reduce the extrapolation but can substantially deflate the importance of correlated predictors. For the related goal of visualizing the effects of each predictor when strong predictor correlation is present, accumulated local effects (ALE) plots were recently introduced and have been widely adopted. This paper presents a new VIM approach based on ALE concepts that avoids both the extrapolation and the VIM deflation problems when predictors are correlated. We contrast, both theoretically and numerically, ALE VIMs with Shapley and permutation VIMs. Our results indicate that ALE VIMs produce similar variable importance rankings as Shapley and permutation VIMs when predictor correlations are mild and more reliable rankings when correlations are strong. An additional advantage is that ALE VIMs are far less computationally expensive.
\end{abstract}

\medskip

\keywords{Variable importance; Shapley values; Accumulated local effects; Model interpretability; Model explainability}

\section{Introduction}\label{Section1Introduction}
\par
With the increasing availability of large data sets and computational resources, black box supervised learning models such as neural networks, boosted trees, random forests, and support vector machines are ubiquitously used to capture complex predictive relationships. While black box models can have much higher accuracy than traditional linear and logistic regression, they are more difficult to interpret in terms of understanding the effects of the predictor variables (predictors for short) on the predicted response. Understanding the predictor effects is crucial for many reasons, e.g., for scientific discovery, trustworthiness, model refinement, and model transparency, in general. 

\par
To address interpretability, various methods have been proposed in recent years. One important class of methods is variable importance measures (VIMs), which are numeric scores that quantify the relative strength of the dependence of the predicted response on predictors. VIMs can be broadly categorized into global versus local and/or model-agnostic versus model-specific (\cite{Ribeiro2016model}). Global VIMs quantify the effect of a predictor over the entire predictor space, whereas local VIMs quantify the effect of a predictor at a single point in the predictor space, e.g., for a specific case of interest. Model-specific VIMs are developed and used for one particular class of supervised learning models, whereas model-agnostic VIMs are more generically applicable to a broad class of models. The focus of this paper is global, model-agnostic VIMs.

\par
Suppose that we have a predictive model of the form $\hat{Y}=f\left(\mathbf{X}\right)$ that was fit to a set of training data, where $Y$ is a random scalar response variable, $\mathbf{X}=\left(X_1,X_2,\ldots,X_d\right)^T$ is a random vector of $d$ predictors, and the function $f\left(\cdot\right)$ is the fitted supervised learning model. For an index set $J\subseteq\left\{1, 2, \ldots,d\right\}$, let $\mathbf{X}_J=(X_k:k\in J)$ denote the subset of predictors with indices included in $J$. To assess the importance of a particular predictor, most existing VIMs require some form of random sampling of subsets $\mathbf{X}_J$ of predictors from either their marginal distribution or their conditional distribution given the other predictors. When predictors are highly correlated, which is usually the case in supervised learning, VIMs based on marginal and conditional distributions often give drastically different results and are each subject to serious shortcomings that a number of recent works have brought to light, e.g., \cite{frye2020asymmetric}, \cite{janzing2020feature}, \cite{kumar2020problems}, \cite{sundararajan2020many}, and \cite{hooker2021}.

\par
In particular, \cite{hooker2021} provides an illuminating discussion contrasting permutation VIMs (\cite{breiman2001random}) with conditional permutation VIMs (\cite{strobl2008conditional}; \cite{fisher2019all}). When estimating the importance of $X_j$, the former samples (via permutation or simulation) $X_j$ from its marginal distribution, while the latter samples $X_j$ from its conditional distribution, which is closely related to refitting $f\left(\cdot\right)$ while excluding the variable of interest and then assessing the extent to which predictive accuracy declines. A major shortcoming of permutation VIMs is that sampling from the marginal distribution of $X_j$ breaks the inherent correlation structure of the predictors and requires extrapolation of $f\left(\mathbf{X}\right)$ to $\mathbf{X}$ values that are far outside the training data envelope, which can render the VIMs unreliable. On the other hand, conditional permutation VIMs avoid extrapolation by design but substantially underestimate the effect (e.g., relative to the coefficients of each $X_j$ if $f\left(\mathbf{X}\right)$ is a linear regression model) of important predictors when they are highly correlated with other predictors, for much the same reason that leaving out a highly correlated predictor typically does not substantially worsen the predictive accuracy. \cite{hooker2021} refers to this as the ``compensation effect" and suggests a way to correct for it primarily for linear regression models. See \cite{hooker2021} and Section \ref{Section3ReviewofVIMs} of this paper for more detailed discussions of these phenomena. 

\par
\cite{hooker2021} recommends conditional permutation VIMs over permutation VIMs, perhaps because they view extrapolation as more concerning than underestimation of importance. Our view is that underestimation of importance is just as concerning, for reasons discussed in Section \ref{Section3ReviewofVIMs}. The main contribution of this paper is to introduce a new set of global, model-agnostic VIMs that not only avoid both of these problems, but that are also computationally inexpensive and have the desirable properties discussed in Section \ref{Section2DesirableProperties}. Many of the most popular existing VIMs (e.g., permutation and conditional permutation VIMs and Shapley values) are very computationally expensive for complex models fit to large data sets. 

\par
Our new VIMs are based on concepts related to accumulated local effects (ALE) plots (\cite{Apley2020}), which are an alternative to partial dependence (PD) plots (\cite{Friedman2001}) for visualizing the global main effects and low-order interaction effects of predictors. In a main effect ALE plot for a predictor $X_j$, \textit{local effect} refers to the effect of a small change in $X_j$ on the predicted response (i.e., a partial derivative or finite difference of $f\left(\mathbf{X}\right))$ and \textit{accumulation} refers to integrating or summing the local effects along paths traversing the predictor space. Let $\mathbf{X}_{\backslash j}$ denote the set of predictors $\mathbf{X}$ excluding $X_j$. For each value of $X_j$, main effect ALE plots first average the local effects with respect to the conditional distribution of $\mathbf{X}_{\backslash j}|X_j$ and then accumulate these averaged local effects across the sample space of $X_j$ to produce a global function of $X_j$ that can be plotted and visualized.  ALE plots were designed to avoid the extrapolation problem in PD plots and are also much less computationally expensive. 

\par
Our new VIMs, which we refer to as path ALE (PALE) VIMs, compute first-order local effects as in ALE plots. However, instead of averaging the local effects (with respect to the conditional distribution of $\mathbf{X}_{\backslash j}|X_j$) prior to accumulating to produce a single function of $X_j$, we accumulate the individual local effects along an appropriately defined collection of paths in the predictor space to produce a collection of functions of $X_j$. The PALE VIMs are essentially the variance of this collection of functions, where the variance is with respect to both $X_j$ and to the functions being randomly drawn from the collection. In addition to avoiding the extrapolation and underestimation problems, the PALE VIMs can be orders of magnitude faster than popular existing VIMs and possess other desirable properties discussed in Section \ref{Section2DesirableProperties}.

\par
The use of accumulated local effects in the PALE VIMs may on the surface sound similar to the integrated gradient (IG) approach of \cite{sundararajan2017axiomatic}, which is a local VIM that quantifies the importance of a predictor $X_j$ on an individual case having predictors $\mathbf{X}=\mathbf{x}$. The IG approach specifies a baseline case with predictors $\mathbf{X}=\mathbf{x'}$ (e.g., the mean of $\mathbf{X}$ for tabular data or a black image for images), and then takes the VIM of $X_j$ to be the integral of the partial derivative of $f\left(\mathbf{X}\right)$ with respect to $X_j$ along a straight-line path from $\mathbf{x'}$ to $\mathbf{x}$. Our approach of defining a global PALE VIM as the variance of a collection of functions of $X_j$, as well as the nature of the paths over which we accumulate the local effects, are fundamentally different than the IG VIM. 

\par
The format of the remainder of the paper is as follows. In Section \ref{Section2DesirableProperties}, we briefly discuss properties that we believe are desirable for global, model-agnostic VIMs to possess. In Section \ref{Section3ReviewofVIMs}, we review some popular existing VIMs and discuss their drawbacks that motivate our PALE VIMs. In Section \ref{Section4VIMsALE}, we provide background on ALE plots and discuss simple VIMs based on them that capture main effect and second-order interaction effect importance of predictors. In Section \ref{Section5PALEVIMs}, we extend certain ALE plot concepts to develop the PALE VIMs that implicitly consider interaction effects of all orders and are therefore total effect VIMs, and we discuss various theoretical properties that they possess. In Section \ref{Section6Examples}, using simulated and real data examples, we illustrate the PALE VIMs and numerically compare them with existing VIMs. We find that when the fitted model $f\left(\mathbf{X}\right)$ is able to extrapolate more reliably, the PALE VIMs produce relative rankings of predictor importance that agree closely with marginal permutation and Shapley VIMs at a fraction of the computational expense. But when $f\left(\mathbf{X}\right)$ extrapolates poorly, the PALE VIMs more reliably agree with reasonable definitions of the ground truth and, when training sample size is smaller, are more consistent across replicates. Section \ref{Section7Conclusions} concludes the paper.

\section{What Properties Should Global, Model-Agnostic VIMs Possess?}\label{Section2DesirableProperties}
\par
In this section we discuss various properties that we believe global, model-agnostic VIMs should possess, recognizing that some of the properties are debatable. We first introduce notation and conventions that we use throughout the paper. Suppose the supervised learning model $f\left(\cdot\right):\mathbb{R}^d\rightarrow\mathbb{R}$ is fit to a training sample $D=\{y_i, \mathbf{x}_i=\left(x_{i,1},x_{i,2},\ldots,x_{i,d}\right):i=1,2,\ldots,n\}$ comprised of $n$ observations (typically treated as i.i.d.) drawn from the joint distribution of the response variable $Y$ and the predictors $\mathbf{X}$. In the regression setting, $f\left(\mathbf{x}\right)$ approximates $\mathbb{E}\left[Y\middle|\mathbf{X}=\mathbf{x}\right]$. For classification, $f\left(\mathbf{x}\right)$ approximates $\mathbb{P}(Y=k|\mathbf{X}=\mathbf{x})$ or $\log(\frac{\mathbb{P}(Y=k|\mathbf{X}=\mathbf{x})}{\mathbb{P}(Y\neq k|\mathbf{X}=\mathbf{x})})$, where the event $\{Y = k\}$ represents the response belonging to a particular class $k$ of interest. Throughout the paper, we use uppercase to denote random variables and lowercase to denote specific values of random variables. For a predictor index set $J\subseteq \{1, 2, \ldots,d\}$, let $\mathbf{X}_{\backslash J}=(X_k:k\notin J)$ denote the subset of predictors with indices not in $J$, and let $\mathbf{x}_{i,J}$ and $\mathbf{x}_{i,\backslash J} (i=1,2,\ldots,n)$ denote the $i$th observation of $\mathbf{X}_J$ and $\mathbf{X}_{\backslash J}$, respectively. We use $p(\cdot)$ to denote the joint probability density (or mass, for discrete numerical predictors) function of $\mathbf{X}$ and notation such as $p_J(\cdot)$ and $p_{\backslash J|J}(\cdot|\cdot)$ to denote the marginal and conditional densities of subsets of $\mathbf{X}$ corresponding to the subscripts. We treat the model $f\left(\mathbf{x}\right)$ as a given, fixed function that one wishes to analyze/interpret via the VIMs, as opposed to treating it as a random function and considering its uncertainty due to estimation error during training. Thus, in expressions like $\mathrm{Var}\left[f(\mathbf{X})\right]$, the source of randomness is $\mathbf{X}\sim p(\mathbf{x})$.

\par
For each predictor $X_j$, let ${VIM}_{j,M}$ and ${VIM}_{j,T}$ respectively denote its main effect (measuring only the “average” effect of $X_j$ without regard to any of its interactions with other variables) and total effect (including the main effect of $X_j$, as well as all interaction effects in which $X_j$ is involved) importance measures. Our use of the term “total effect” is as in the global sensitivity analysis literature, which is different than in the causal inference literature. We believe the following are all desirable properties for VIMs to possess:
\begin{enumerate}[(i)]
    \item \textit{Reasonable computational expense}: Their computational expense should be reasonable even for large data sets and complex functions $f\left(\mathbf{x}\right)$.
    
    \item \textit{Avoidance of extrapolation}: They should be reliable with correlated predictors (because correlation is the norm in large observational data sets) without requiring extrapolation of $f\left(\mathbf{x}\right)$ outside the training data envelope.
    
    \item \label{property: doesnotappear} \textit{Zero importance for absent predictors}:  If a predictor $x_j$ does not appear anywhere in $f\left(\mathbf{x}\right)$ (i.e., for each fixed $\mathbf{x}_{\backslash j}$, $f\left(\mathbf{x}_{\backslash j},x_j\right)$ assumes the same value for all $x_j$), then we should have ${VIM}_{j,T}={VIM}_{j,M}=0$. 

    \item \label{property: total>main}
    \textit{Total effects that can be compared to main effects}:  There should be both main effect and total effect VIMs that can be compared to help assess the strengths of interactions. In particular, ${VIM}_{j,T}\geq{VIM}_{j,M}$ should always hold, with equality if and only if $f\left(\mathbf{x}\right)$ is additive in $x_j$, i.e., if and only if we can write the model as $f\left(\mathbf{x}\right)=f_j\left(x_j\right)+f_{\backslash j}\left(\mathbf{x}_{\backslash j}\right)$ for some functions $f_j\left(x_j\right)$ and $f_{\backslash j}\left(\mathbf{x}_{\backslash j}\right)$.  ${VIM}_{j,T}\gg{VIM}_{j,M}$ should indicate that $X_j$ interacts strongly with some other predictors. 

    \item \label{property: additive recovery} \textit{Additive recovery}: If $f\left(\mathbf{x}\right)=f_j\left(x_j\right)+f_{\backslash j}\left(\mathbf{x}_{\backslash j}\right)$ is additive in $x_j$, then ${VIM}_{j,T}={VIM}_{j,M}$ should depend only on $f_j\left(X_j\right)$, e.g., ${VIM}_{j,T}={VIM}_{j,M}=\mathrm{Var}\left[f_j\left(X_j\right)\right]$. 
        
    \item \textit{Avoidance of surrogate models}: They should be computable from only the fitted model $f\left(\mathbf{x}\right)$ and the empirical distribution of the predictors $\{\mathbf{x}_i: i=1,2,\ldots,n\}$ over the training data, if the goal is to assign importance to predictors based on how they influence predictions from the model of interest $f\left(\mathbf{x}\right)$. In particular, they should avoid fitting surrogate models to $f\left(\mathbf{x}\right)$ or refitting $f\left(\mathbf{x}\right)$ to the training data with predictors omitted or permuted, which can be computationally expensive and depend on the details of how the additional models are fit.
    
    \item \textit{Avoidance of modeling  $p(\mathbf{x})$}: They should not require generating data directly from $p(\mathbf{x})$ or \\$p_{j|\backslash j}(x_j|\mathbf{x}_{\backslash j})$, since these distributions are unknown in practice and difficult to estimate for high-dimensional predictors.
    
\end{enumerate}

\par
The VIMs that we develop in this paper have all of these properties. Our view that these properties are desirable assumes that the objective is to interpret a given function $f\left(\mathbf{x}\right)$ that represents a fitted model, in terms of how the predictors \textit{mathematically} affect the predictions produced by $f\left(\mathbf{x}\right)$. If $f\left(\mathbf{x}\right)$ is a poor model because of improper modeling practice (e.g., poor hyperparameter tuning) or poor quality data, then VIMs based on $f\left(\mathbf{x}\right)$ will inevitably be misleading, but we view this as a problem with $f\left(\mathbf{x}\right)$ itself and not the VIM method. The most we can ask of a VIM method is to accurately interpret the $f\left(\mathbf{x}\right)$ that we have available. Similarly, if $f\left(\mathbf{x}\right)$ is fit to observational data, one should be careful interpreting the VIMs as representing causal effects of the predictors. To illustrate, a simple causality example discussed in \cite{pearl2018book} concerns the relationship between scurvy ($Y$) in sailors and citrus consumption ($X_1$). Regressing $Y$ onto only $X_1$ might lead one to conclude that citrus is what prevents scurvy, but it was unfortunately observed that giving sailors boiled citrus juice did not. In this case the fitted model does not capture full causality relationships because of omitted variable bias. Namely, vitamin C, which is what causally prevents scurvy but is degraded by boiling, was omitted from the regression. If a study was conducted in which vitamin C consumption was also measured and included as a predictor in the regression, interpretation of the resulting fitted $f\left(\mathbf{x}\right)$ would likely have led to a more accurate understanding of factors that causally prevent scurvy. 

\par
Addressing such causality issues is complex and beyond the scope of this paper, as it would typically require fitting a different model after collecting more informative data (e.g., experimental) and/or incorporating physical knowledge of causality flows. See, for example, \cite{pearl2016causal} for causal inference considerations and methods. Our goal of interpreting the mathematical effects of each predictor on a given function $f\left(\mathbf{x}\right)$ translates to interpreting the “direct effects” (using causal inference terminology) of the predictors. In particular, property (\ref{property: doesnotappear}) should hold in this goal, whereas in causal inference it would not necessarily hold. Identifying the variables with largest direct effects on $f\left(\mathbf{x}\right)$ is useful in and of itself, even if only to inform follow-up studies to identify whether variables with large effect are causally influenced by other variables. 

\par
In the above list, we have omitted the property that $\mathrm{Var}\left[f\left(\mathbf{X}\right)\right]=\sum_{j=1}^{d}{VIM}_{j,T}$, i.e., that portions of $\mathrm{Var}\left[f\left(\mathbf{X}\right)\right]$ can be allocated across the $d$ predictors, a property that also seems desirable on the surface. However, it has some negative side effects and directly conflicts with property (\ref{property: additive recovery}). To see the conflict, consider the linear regression model $f\left(\mathbf{X}\right)=\beta_1 X_1+\beta_2 X_2$, where $\left(X_1,X_2\right)$ are standard bivariate normal random variables having correlation $\rho$. Property (\ref{property: additive recovery}) would result in ${VIM}_{j,T}=\mathrm{Var}\left[\beta_j X_j\right]=\beta_j^2$, which seems desirable and is consistent with the standard way to interpret the importance of variables in linear regression models and in additive models of the form $f\left(\mathbf{x}\right)=\sum_{j=1}^{d}{f_j\left(x_j\right)}$, for which the importance of $X_j$ is typically taken to be $\mathrm{Var}\left[f_j\left(X_j\right)\right]$. In contrast, if we required the VIMs to sum to $\mathrm{Var}\left[f\left(\mathbf{X}\right)\right]$, then the VIM for $X_1$ would depend heavily on $\rho$ and on $\beta_2$, even though the direct effect of a unit change in $X_1$ on $f\left(\mathbf{X}\right)$ remains the same value $\beta_1$ for any $\rho$ and $\beta_2$. This would be especially problematic if the correlation between $X_1$ and $X_2$ that occurred in the training data could be broken to some extent in the future by more independently controlling $X_1$ and $X_2$. For example, if $\beta_1 = \beta_2$ and $\rho$ is large and negative over the training data, requiring the VIMs to sum to $\mathrm{Var}\left[f\left(\mathbf{X}\right)\right]$ would result in $VIM_{1,T} \approx VIM_{2,T} \approx 0$ and imply that neither $X_1$ nor $X_2$ influence the response when they both do, especially if the correlation can be broken. In light of the above, we believe property (\ref{property: additive recovery}) is typically more desirable than the VIMs summing to $\mathrm{Var}\left[f\left(\mathbf{X}\right)\right]$, although this could depend on the application and how one intends to use the VIMs. It should be noted that neither of the two most commonly used VIMs (permutation-based or the most popular Shapley-based implementations, see Section \ref{Section3ReviewofVIMs}) have the property that the VIMs sum to $\mathrm{Var}\left[f\left(\mathbf{X}\right)\right]$.

\section{Review of Various Existing Global, Model-Agnostic VIMs}\label{Section3ReviewofVIMs}
\par 
\textbf{\textit{Global Sensitivity Analysis indices}}: Global sensitivity analysis (GSA; see \cite{saltelli2008global} for a review) indices are closely related to VIMs. They are used to assess the importance of the input variables $\mathbf{X}$ when $f\left(\mathbf{X}\right)$ is some black box function, typically a surrogate model fit to the response surface in a computationally expensive computer simulation model.  We focus on variance-based GSA indices, since derivative-based GSA indices (e.g.,  \cite{morris1991factorial}, \cite{campolongo2007effective}, \cite{sobolkucherenko2009derivative}) are problematic as VIMs for supervised learning models for a number of reasons. Standard variance-based GSA indices usually assume the inputs are independent, since inputs can often be independently controlled in a simulation model. In this case, GSA indices are closely related to the Sobol decomposition (\cite{sobol1993sensitivity})
\begin{equation} \label{eq: Sobol decomposition}
f\left(\mathbf{x}\right)=f_0+\sum_{J\subseteq\left\{1,2,\ldots,d\right\}}{f_J\left(\mathbf{x}_J\right)}
\end{equation}
of $f\left(\cdot\right)$ into a set of uncorrelated component functions of all subsets of variables of size ranging from $1,2,\ldots,d$. As a result, $\mathrm{Var}\left[f\left(\mathbf{X}\right)\right]=\sum_{J\subseteq\left\{1,2,\ldots,d\right\}}\mathrm{Var}\left[f_J\left(\mathbf{X}_J\right)\right]$. The two most common variance-based indices related to \eqref{eq: Sobol decomposition} are the main effect and total effect indices for $X_j$, defined as (\cite{homma1996importance})
\begin{equation} \label{eq: GS var main}
{GS}_{j,M}=\mathrm{Var}\left[\mathbb{E}\left[f\left(X_j,\mathbf{X}_{\backslash j}\right)|X_j\right]\right],
\end{equation}
and 
\begin{equation}\label{eq: GS var total}
{GS}_{j,T}=\mathbb{E}\left[\mathrm{Var}\left[f\left(X_j,\mathbf{X}_{\backslash j}\right)|\mathbf{X}_{\backslash j}\right]\right].
\end{equation}
Because of the independence assumption and other properties of the Sobol decomposition, they have the properties ${GS}_{j,M}=\mathrm{Var}\left[f_j\left(X_j\right)\right]$ and $\sum_{j=1}^{d}{GS}_{j,T}\geq \mathrm{Var}\left[f\left(\mathbf{X}\right)\right]\geq\sum_{j=1}^{d}{GS}_{j,M}$, where the inequalities are replaced by equalities when $f\left(\cdot\right)$ is additive in all inputs. 

\par
The decomposition \eqref{eq: Sobol decomposition} applies only to the case of independent inputs, in which case ${GS}_{j,T}$ is an appealing measure of the total importance of $X_j$. For dependent inputs, the situation is far more complex. \cite{Hooker2007} generalized \eqref{eq: Sobol decomposition} to dependent inputs with the primary goal of avoiding extrapolation when computing estimates of the component functions, and a number of follow-up works (e.g., \cite{Li2012} and \cite{Chastaing2012}) proposed modified versions and/or developed computational approaches. The approaches are cumbersome and computationally expensive to implement (e.g., requiring specialized surrogate models for $f(\cdot)$ and explicit computation of many component functions), and, to the best of our knowledge, are rarely used for VIMs for supervised learning models fit to observational data.   

\par
It should be noted that \eqref{eq: GS var main} and \eqref{eq: GS var total} have serious flaws when the predictors are highly correlated. Regarding \eqref{eq: GS var main}, $\mathbb{E}\left[f\left(X_j,\mathbf{X}_{\backslash j}\right)|X_j=x_j\right]$ as a function of $x_j$ represents the effects of both $X_j$ and any components of $\mathbf{X}_{\backslash j}$ that are correlated with $X_j$, akin to the omitted variable bias problem in regression. Regarding \eqref{eq: GS var total}, ${GS}_{j,T}$ substantially underestimates the importance of $X_j$ when it is highly correlated with other predictors, because $\mathrm{Var}\left[X_j|\mathbf{X}_{\backslash j}\right]$ and thus $\mathrm{Var}\left[f\left(X_j,\mathbf{X}_{\backslash j}\right)|\mathbf{X}_{\backslash j}\right]$ are too small to reflect the true importance of $X_j$. We demonstrate this in Section \ref{Section6.1TheoreticalExample} for the linear model $f\left(\mathbf{X}\right)=\beta_1 X_1+\beta_2 X_2+\beta_3 X_3$ with bivariate standard normal $\left(X_1, X_2\right)$ having correlation $\rho$ and independent standard normal $X_3$, in which case both ${GS}_{1,T}=\beta_1^2\left(1-\rho^2\right)$ and ${GS}_{2,T}=\beta_2^2\left(1-\rho^2\right)$ decay to zero as $\rho\rightarrow\pm 1$. This could misleadingly imply that $X_3$ has higher importance than either $X_1$ or $X_2$ when $\left|\beta_3\right|\ll\left|\beta_1\right|\approx\left|\beta_2\right|$.

\par
\textbf{\textit{Permutation VIMs}}: \cite{breiman2001random} first proposed the permutation VIM in the specific context of random forest models. Suppose that $\mathrm{ntree}$ bootstrap samples $\{D_1,D_2, \ldots,D_{\mathrm{ntree}}\}$ are drawn from $D$, and let $h_k\left(\cdot\right)$ denote the tree fit to $D_k$. The random forest model is $f\left(\mathbf{x}\right)=\frac{1}{\mathrm{ntree}}\sum_{k=1}^{\mathrm{ntree}}{h_k(\mathbf{x})}$ for regression or the majority vote for classification. The permutation VIM for $X_j$ is the mean decrease in prediction accuracy, measured by the increase in out-of-bag (OOB) error, when values of $X_j$ are randomly permuted over their marginal distribution in the OOB samples. Model-agnostic versions have also been proposed (see, e.g., \cite{fisher2019all}), although for computational reasons the OOB error is replaced by the training error after permutation. 

\par
\begin{figure}[!ht]
\centering
\includegraphics[width=\textwidth]{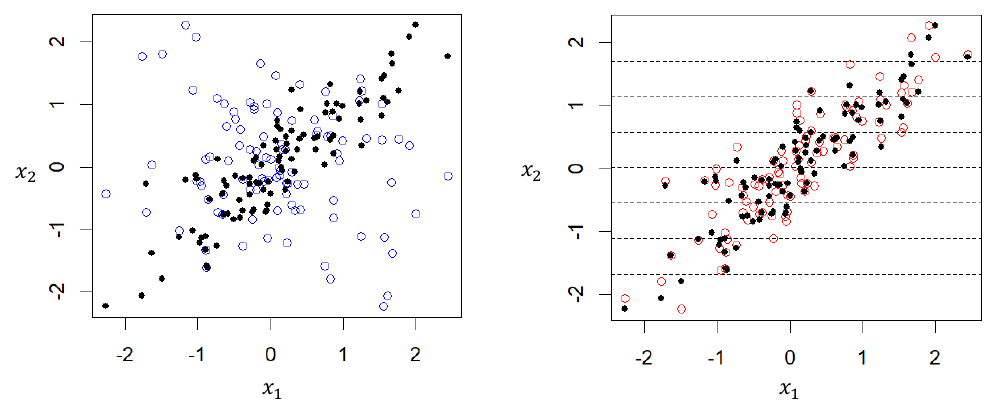}
\caption{Illustration of problems with permutation (left panel) and conditional permutation (right panel) VIMs when predictors are highly correlated. The bullets in both plots are scatter plots of $\left(x_1,x_2\right)$ for the training data. The blue circles in the left plot are the same but with the $x_1$ values randomly permuted, resulting in many values falling far outside the training data envelope into regions (e.g., in the upper-left and lower-right corners) where the extrapolated $f\left(\cdot\right)$ would typically be unreliable. The red circles in the right plot are a conditionally permuted sample (permutation of $x_1$ occurs only within the $x_2$ bins), which avoids substantial extrapolation but results in underestimation of the importance of $X_1$.}
\label{fig: permuted bivariate normal data}
\end{figure}

In addition to having prohibitive computational expense for general $f\left(\cdot\right)$ when OOB or cross-validation permutation error is used, the permutation VIM is subject to the extrapolation problem when the predictors are highly correlated. The reason is that randomly permuting a single predictor and leaving the others unchanged results in a vector of predictors that no longer follows the joint distribution $p(\cdot)$ from which the training samples were drawn. The left panel of Fig. \ref{fig: permuted bivariate normal data} illustrates this for the case $d=2$, in which we plot $n=100$ training observations $\left\{\left(x_{i,1},x_{i,2}\right):i=1, 2,\ldots,100\right\}$ drawn from a standard bivariate normal distribution with correlation coefficient $\rho=0.9$. We also show the same training sample but with the $X_1$ values randomly permuted. Many of the permuted values of $\left(x_1,x_2\right)$ fall far outside the envelope of the training data, which would require substantial extrapolation when these values are plugged into $f\left(\cdot\right)$. This extrapolation problem is well-known and has been discussed in many recent works, e.g., \cite{hooker2021}.

\par
\textbf{\textit{Conditional permutation VIMs}}: To avoid problems with the original permutation VIMs when the predictors are highly correlated, \cite{strobl2008conditional} first proposed the conditional permutation VIM, also in the context of random forests. Like the original permutation VIM, the conditional permutation VIM has been extended in various works, including several model-agnostic versions in \cite{fisher2019all}. To compute the VIM for $X_j$, the algorithm first finds a set of predictor indices $L\subseteq\{1,2,\ldots,d\}\backslash\{j\}$ such that $\mathbf{X}_L$ are highly correlated with $X_j$. For each tree $h_k\left(\cdot\right)$, bins are formed in the $\mathbf{X}_L$ space consistent with the $\mathbf{X}_L$ split point values in the tree partition. The $x_{i,j}$ values are permuted only within each bin to encourage the permuted values to follow approximately the correct conditional distribution $p_{j|L}(\cdot|\cdot)$. We illustrate this in the right panel of Fig. \ref{fig: permuted bivariate normal data} for the same example. In the lower-dimensional setting, if the bins are sufficiently narrow, this effectively avoids substantial extrapolation. However, when $L$ is large the bins in the $\mathbf{X}_L$ space must also be large to contain enough observations to permute, in which case conditional permutation may still require substantial extrapolation. 

\par
Perhaps more importantly, conditional permutation VIMs can substantially underestimate variable importance when predictors are highly correlated, for the same reason that the GSA total sensitivity index \eqref{eq: GS var total} does. Namely, $\mathrm{Var}\left[X_j|\mathbf{X}_{\backslash j}\right]$ will be small if there is high correlation (or other nonlinear forms of dependence) between $X_j$ and other predictors, in which case $\mathrm{Var}\left[f\left(X_j,\mathbf{X}_{\backslash j}\right)|\mathbf{X}_{\backslash j}\right]$ will be small for smooth $f\left(\cdot\right)$. This implies that permuting $X_j$ within individual bins will have little effect on the prediction accuracy, since selecting a bin within which $X_j$ will be permuted represents conditioning on $\mathbf{X}_{\backslash j}$. We demonstrate this in Section \ref{Section6.1TheoreticalExample} for the same linear model $f\left(\mathbf{X}\right)=\beta_1 X_1+\beta_2 X_2+\beta_3 X_3$ with correlated $\left(X_1, X_2\right)$, in which case the theoretical limiting form of the conditional permutation VIMs for $X_1$ and $X_2$ are the same as the variance-based GSA total effect indices and decay to zero as $\rho\rightarrow\pm 1$. \cite{hooker2021} discusses this further and shows that the conditional permutation VIM estimates the same target quantity as dropping the correlated variable of interest, refitting the model, and taking its importance to be the extent to which predictive accuracy declines. 

\par
\textbf{\textit{Shapley value-based VIMs}}: Shapley values (\cite{shapley1953value}) from cooperative game theory have been adapted to assign variable importance to predictors (e.g., \cite{owen2014sobol}; \cite{vstrumbelj2014}). One popular version is a local method that attempts to distribute $f\left(\mathbf{x}\right)-\mathbb{E}[f\left(\mathbf{X}\right)]$ for each case $\mathbf{x}=\left(x_1,x_2,\ldots,x_d\right)$ of interest to contributions from each predictor. The contribution of $x_j$ is defined as (see, e.g., \cite{vstrumbelj2014}, \cite{lundberg2017unified})
\begin{equation}\label{eq: Shapley value}
\phi_j\left(\mathbf{x}\right)=\sum_{u\subseteq\{1,2,\ldots,d\}\backslash \{j\}}\frac{\left|u\right|!\left(d-\left|u\right|-1\right)!}{d!}\left[f_{u\cup{j}}\left(\mathbf{x}_{u\cup{j}}\right)-f_u\left(\mathbf{x}_u\right)\right],
\end{equation}
where $f_u\left(\mathbf{x}_u\right)$ is some value function chosen for the predictors $\mathbf{X}_u$. To assess global importance for $X_j$, $\phi_j\left(\mathbf{x}_i\right)$ (or its absolute value or square) can be aggregated across a representative subset of observations $\mathbf{x}_i$ or the entire training set, although there are serious problems with this (see the example below).

\par
When choosing an appropriate value function $f_u\left(\mathbf{x}_u\right)$ in \eqref{eq: Shapley value}, the same permutation VIM debate on using marginal versus conditional distributions arises with similar tradeoffs. \cite{lundberg2017unified} consider $f_u\left(\mathbf{x}_u\right) = \mathbb{E}\left[f(\mathbf{x}_u,\mathbf{X}_{\backslash u})|\mathbf{X}_u=\mathbf{x}_u\right]$ but focus on $f_u\left(\mathbf{x}_u\right)\approx \mathbb{E}\left[f\left(\mathbf{x}_u,\mathbf{X}_{\backslash u}\right)\right]$ in their computational procedure, which samples from the marginal joint distribution of $\mathbf{X}_{\backslash u}$. As noted in many works (e.g., \cite{frye2021shapley}, \cite{sundararajan2020many}, \cite{chen2020true}, \cite{hooker2021}), estimating $f_u\left(\mathbf{x}_u\right)\approx \mathbb{E}\left[f\left(\mathbf{x}_u,\mathbf{X}_{\backslash u}\right)\right]$ requires unreliable extrapolation of $f(\cdot)$ outside the training data envelope. To avoid this problem, a number of authors (e.g., \cite{frye2021shapley}; \cite{aas2021explaining}) have developed methods to estimate $\mathbb{E}\left[f(\mathbf{x}_u,\mathbf{X}_{\backslash u})|\mathbf{X}_u=\mathbf{x}_u\right]$ by sampling from the conditional distribution of $\mathbf{X}_{\backslash u}|\mathbf{X}_u=\mathbf{x}_u$.

\par
However, as noted by  \cite{sundararajan2020many}, \cite{janzing2020feature}, and others, using $f_u\left(\mathbf{x}_u\right)=\mathbb{E}\left[f(\mathbf{x}_u,\mathbf{X}_{\backslash u})|\mathbf{X}_u=\mathbf{x}_u\right]$ can assign non-zero importance to predictors not used by the model. It can also underestimate the importance of important predictors. We also demonstrate these drawbacks in Section \ref{Section6.1TheoreticalExample} for the same linear model $f\left(\mathbf{X}\right)=\beta_1 X_1+\beta_2 X_2+\beta_3 X_3$ with correlated $\left(X_1, X_2\right)$. In particular, when $\beta_2=0$ the global Shapley VIMs for $X_1$ and $X_2$ obtained by averaging the squares of the local values are $\left(1-\frac{3\rho^2}{4}\right)\beta_1^2$ and $\frac{\rho^2\beta_1^2}{4}$, respectively (see Section \ref{Section6.1TheoreticalExample}). Even though $X_2$ is completely absent from the model and therefore has no direct effect, its Shapley VIM approaches the VIM for $X_1$ as $\left|\rho\right|\rightarrow1$. This seems like a severe drawback of Shapley VIMs that use conditional distributions in $f_u\left(\mathbf{x}_u\right)$. Although Shapley VIMs using marginal distributions in $f_u\left(\mathbf{x}_u\right)$ would give more “correct” VIMs of $\beta_1^2$ and $0$ for $X_1$ and $X_2$, respectively, in this example, they require severe extrapolation when $|\rho|$ is large. 

\par
One common justification for local Shapley VIMs, referred to as “Property 1: Local accuracy” in \cite{lundberg2017unified} and “Axiom 1: Efficiency” in \cite{frye2020asymmetric}, implies that $\sum_{j=1}^{d}{\phi_j\left(\mathbf{x}\right)}=f\left(\mathbf{x}\right)-\mathbb{E}\left[f\left(\mathbf{X}\right)\right]$, i.e., that the difference between the prediction $f\left(\mathbf{x}\right)$ and the average prediction can be decomposed into contributions from each of the $d$ predictors. If we extend the local Shapley VIMs to global VIMs by averaging them over all training data so that the global VIM for $X_j$ is $\frac{1}{n}\sum_{i=1}^{n}{\phi_j\left(\mathbf{x}_i\right)}\approx \mathbb{E}\left[\phi_j\left(\mathbf{X}\right)\right]$, the analogous decomposition only implies that the global VIMs sum to zero, which is not particularly useful (in the preceding simple example, this would produce global VIMs for $X_1$ and $X_2$ that are both identically zero). A more common global Shapley VIM averages the absolute values or squares of the local VIMs, e.g., $\frac{1}{n}\sum_{i=1}^{n}\left|\phi_j\left(\mathbf{x}_i\right)\right|$, but this global VIM does not inherit any analogous decomposition property. One global Shapley VIM that does have a decomposition property for general correlated $\mathbf{X}$ was proposed by \cite{song2016shapley} and uses value function $\mathbb{E}\left[\mathrm{Var}\left[f(\mathbf{X}_u,\mathbf{X}_{\backslash u})|\mathbf{X}_{\backslash u}\right]\right]$ for variable index subset $u$ in a global version of \eqref{eq: Shapley value}. By a fundamental property of Shapley values, this results in the VIMs for all variables summing to $\mathrm{Var}\left[f(\mathbf{X})\right]$. However, as discussed in Section \ref{Section2DesirableProperties}, this has the undesirable consequences of violating properties (\ref{property: total>main}) and (\ref{property: additive recovery}). Its computational expense is also substantially higher than for the more common Shapley VIMs discussed above (which are themselves computationally expensive) because estimating $\mathbb{E}\left[\mathrm{Var}\left[f(\mathbf{X}_u,\mathbf{X}_{\backslash u})|\mathbf{X}_{\backslash u}\right]\right]$ for each of the many subsets $u$ requires two-level nested sampling of $\mathbf{X}_u$ and $\mathbf{X}_{\backslash u}$. 

\par
To summarize, the primary existing model-agnostic global VIMs (permutation and Shapley VIMs) suffer from the same difficult tradeoff---sampling from marginal predictor distributions provides more conceptually reasonable VIMs but involves unreliable extrapolation, while sampling from conditional distributions avoids or partially avoids extrapolation but provides VIMs that can be severely misleading, e.g., underestimating importance of important predictors and attributing substantial importance to predictors not used by the model. Moreover, both approaches involve heavy computational expense, whereas the PALE VIMs avoid all of these serious drawbacks of existing VIMs. 

\section{VIMs Based on ALE Plots}\label{Section4VIMsALE}
\par
A task related to assigning a numeric VIM to a predictor $X_j$ is to visualize the global main effect of $X_j$ and the interaction effect of a pair $(X_j,X_l)$ on $f\left(\cdot\right)$. In this section, we review ALE and partial dependence (PD) plots, which are used for these purposes. Although we focus on ALE plots, both methods yield a straightforward and natural measure of main-effect and second-order interaction effects importance, which we discuss in Section \ref{Section4.2ALEMainSecondVIMs}.  In Section \ref{Section5PALEVIMs} we extend certain ALE plot concepts to develop the PALE total effect VIMs that implicitly consider interactions of all orders.

\subsection{Background on ALE Plots}\label{Section4.1BackgroundALE}
\par
To understand the effects of predictors on $f\left(\cdot\right)$, we must do some form of averaging across values of the other predictors. For example, in \cite{Friedman2001}, the partial dependence (PD) main effect of $X_j$ is defined as $\mathbb{E}\left[f\left(x_j,\mathbf{X}_{\backslash j}\right)\right]$ as a function of $x_j$, where the expectation is with respect to the marginal distribution of $\mathbf{X}_{\backslash j}$. The PD interaction effect of $(X_j,X_l)$ is defined analogously, as $\mathbb{E}\left[f\left(x_j,x_l,\mathbf{X}_{\backslash\{j,l\}}\right)\right]$, as a function of $(x_j,x_l)$. ALE plots are interpreted the same way as PD plots but are defined and computed differently to avoid problems with PD plots, namely, computational expense and the need to extrapolate $f\left(\cdot\right)$ outside the data range when plugging in $\left(x_j,\mathbf{X}_{\backslash j}\right)$ values with $\mathbf{X}_{\backslash j}$ drawn from its marginal distribution.

\par
We illustrate for the case that $f\left(\cdot\right)$ is differentiable, although ALE plots apply equally to nondifferentiable $f\left(\cdot\right)$, such as tree-based models, with derivatives replaced by finite differences. Suppose $p$ has compact support $\mathcal{S}$ and the support of $p_j$ is the interval $\mathcal{S}_j=[x_{\min,j}, x_{\max,j}]$ for each $j\in \{1,2,\ldots,d\}$. Let $f^j\left(x_j, \mathbf{x}_{\backslash j}\right)\equiv\frac{\partial f(x_j, \mathbf{x}_{\backslash j})}{\partial x_j}$ denote the first-order partial derivative of $f\left(\mathbf{x}\right)$ with respect to $x_j$. The uncentered ALE main effect function for $X_j$ is defined as, for each $x_j\in\mathcal{S}_j$,
\begin{equation} \label{eq: ALE main uncentered}
g_{j,ALE}\left(x_j\right)= \int_{x_{\min,j}}^{x_j}{\mathbb{E}\left[f^j\left(X_j,\mathbf{X}_{\backslash j}\right)|X_j=z_j\right]dz_j},
\end{equation}
which avoids the extrapolation required in PD plots by using the conditional distribution of $\textbf{X}_{\backslash j}|X_j$. The centered (zero-mean) ALE main effect function of $X_j$, denoted by $f_{j,ALE}\left(x_j\right)$ is defined as
\begin{equation} \label{eq: ALE main centered}
f_{j,ALE}\left(x_j\right)=g_{j, ALE}\left(x_j\right)-\mathbb{E}\left[g_{j, ALE}\left(X_j\right)\right].
\end{equation}

\par
\cite{Apley2020} also defined an ALE second-order interaction function $f_{\{j,l\},ALE}(x_j,x_l)$ for each pair of predictors $(X_j,X_l)$ (we omit the definition and details for brevity), which has the properties $\mathbb{E}\left[f_{\{j,l\},ALE}\left(X_j,X_l\right)\right]=0$ and the ALE main effects of $X_j$ and of $X_l$ on $f_{\{j,l\},ALE}\left(x_j,x_l\right)$ are both zero, so that $f_{\{j,l\}, ALE}\left(x_j,x_l\right)$ captures pure interaction effect between $X_j$ and $X_l$. \cite{Apley2020} showed that ALE effect functions have additive recovery properties analogous to property (\ref{property: additive recovery}): When $f\left(\mathbf{x}\right)=f_j\left(x_j\right)+f_{\backslash j}(\mathbf{x}_{\backslash j})$ is additive in a predictor $x_j$, the ALE main effect function $f_{j,ALE}(x_j)$ is equal to the correct function $f_j\left(x_j\right)$, up to an additive constant. When $x_j$ only has main effect and second-order interaction effects with other predictor variables in $\mathbf{X}_{\backslash j}$, i.e.\ $f\left(\mathbf{x}\right)=f_j\left(x_j\right)+\sum_{l\neq j}{f_{\{j,l\}}(x_j,x_l)}+f_{\backslash j}(\mathbf{x}_{\backslash j})$, the ALE second-order interaction effect $f_{\left\{j,l\right\},ALE}(x_j,x_l)$ is the correct function $f_{\left\{j,l\right\}}\left(x_j, x_l\right)$, up to additive functions of the individual variables. Together,  $\mathbb{E}\left[f(\mathbf{X})\right]+f_{j, ALE}\left(x_j\right)+f_{l, ALE}\left(x_l\right)+f_{\{j,l\}, ALE}\left(x_j, x_l\right)$ constitutes a second-order approximation of the joint effect of $X_j$ and $X_l$, which is visualized by plotting the sample version of this function (discussed below) vs.\ $\left(x_j,x_l\right)$. Likewise, the main effect of $X_j$ can be visualized by plotting the sample version of $\mathbb{E}\left[f(\mathbf{X})\right]+f_{j,ALE}\left(x_j\right)$ vs.\ $x_j$.

\par
Sample versions of the ALE functions are computed using the same training data to which $f(\cdot)$ was fit, by replacing the integrals, derivatives, and expectations in \eqref{eq: ALE main uncentered} and \eqref{eq: ALE main centered} by summations, finite differences, and sample averages, respectively. More specifically, to compute the sample version of the ALE main effect of $X_j$, let $\{N_j\left(k\right)=\left(z_{k-1,j}, z_{k, j}\right]:k=1,2,\ldots,K\}$ be a  partition of the sample range of $\left\{x_{i,j}:i=1,2,\ldots,n\right\}$ into $K$ intervals. For $k=1,2,\ldots, K$, let $n_j(k)$ denote the number of training observations that fall into the $k$th interval $N_j(k)$ so that $\sum_{k=1}^{K}{n_j\left(k\right)=n}$. For a particular value $x$ of $X_j$, let $k_j(x)$ denote the index of the interval into which $x$ falls, i.e.\ $x\in(z_{k_j\left(x\right)-1,j}, z_{k_j\left(x\right),j}]$. A choice of partition that works well in practice is to let $z_{k,j}$ be the $\frac{k}{K}$ quantile of the empirical distribution of $\left\{x_{i, j}:i=1,2,\ldots,n\right\}$ for $k=1,2,\ldots, K$, with $z_{0,j}$ chosen just below the smallest observation. This results in approximately the same number of training observations $n_j\left(k\right)\approx\frac{n}{K}$ in each interval. Throughout the paper, we use this choice of partition. See the left panel of Fig.\ \ref{fig: path motivation 1 additive}, discussed later, for an illustration of the notation. 

\par
The sample versions of \eqref{eq: ALE main uncentered} and \eqref{eq: ALE main centered} are, for each $x\in(z_{0,j},z_{K,j}]$,
\begin{equation} \label{eq: estimator of ALE main uncentered}
{\hat{g}}_{j,ALE}\left(x\right)=\sum_{k=1}^{k_j\left(x\right)}\frac{1}{n_j\left(k\right)}\sum_{\{i:x_{i,j}\in N_j\left(k\right)\}}\left[f\left(z_{k,j}, \mathbf{x}_{i,\backslash j}\right)-f\left(z_{k-1,j},\mathbf{x}_{i,\backslash j}\right)\right]
\end{equation}
and 
\begin{equation} \label{eq: estimator of ALE main centered}
{\hat{f}}_{j,ALE}\left(x\right)={\hat{g}}_{j, ALE}\left(x\right)-\frac{1}{n}\sum_{i=1}^{n}{{\hat{g}}_{j,ALE}\left(x_{i,j}\right)}={\hat{g}}_{j,ALE}\left(x\right)-\frac{1}{n}\sum_{k=1}^{K}{n_j\left(k\right){\hat{g}}_{j,ALE}\left(z_{k,j}\right)}.
\end{equation}

\par
Refer to \cite{Apley2020} for definition and sample versions of ALE second-order effects. 

\subsection{ALE-Based VIMs for Main Effect and Second-Order Interaction Effect Importances}\label{Section4.2ALEMainSecondVIMs}
\par
In this Section we define simple ALE-based measures of main effect and second-order effect importance of predictors. Similar VIMs using PD instead of ALE functions have been previously proposed, e.g., in \cite{friedman2008predictive} and \cite{greenwell2018simple}. For any predictor $X_j$, define its ALE main effect importance ${ALE}_{j,M}$ as the variance of $f_{j,ALE}\left(X_j\right)$ with respect to the marginal distribution of $X_j$, i.e.,
\begin{equation}\label{eq: ALE main VIM}
{ALE}_{j,M}\equiv\mathrm{Var}\left[f_{j,ALE}\left(X_j\right)\right]=\mathbb{E}\left[f_{j,ALE}^2\left(X_j\right)\right].
\end{equation}
Due to the additive recovery property, if $f\left(\mathbf{x}\right)=f_j\left(x_j\right)+f_{\backslash j}(\mathbf{x}_{\backslash j})$, ${ALE}_{j,M}$ is the variance of $f_j\left(X_j\right)$. For a pair of predictors $\left(X_j,X_l\right)$ with $j\neq l$, the importance of its second-order interaction could similarly be defined as $\mathrm{Var}\left(f_{\left\{j,l\right\},ALE}\left(X_j,X_l\right)\right)$. 

\par
For a predictor $X_j$, we can define its importance due to main and second-order interactions as
\begin{equation}\label{eq: ALE second VIM}
\begin{aligned}
{ALE}_{j,2} &\equiv\mathrm{Var}(f_{j,ALE}\left(X_j\right)+\sum_{l\neq j}{f_{\left\{j,l\right\},ALE}\left(X_j, X_l\right)}) \\
&=\mathbb{E}[(f_{j,ALE}\left(X_j\right)+\sum_{l\neq j}{f_{\left\{j,l\right\},ALE}\left(X_j, X_l\right)})^2],
\end{aligned}
\end{equation}
i.e., the variance of the sum of all main effect and second-order interaction effects involving $X_j$. Again, due to additive recovery property, if $f\left(\mathbf{x}\right)=f_j\left(x_j\right)+\sum_{l\neq j}{f_{\left\{j,l\right\}}\left(x_j, x_l\right)}+f_{\backslash j}(\mathbf{x}_{\backslash j})$, ${ALE}_{j,2}$ is the variance of the sum of all terms in $f\left(\textbf{x}
\right)$ involving $X_j$. Sample versions of ${ALE}_{j,M}$ and ${ALE}_{j,2}$ can be obtained by taking the variances of sample versions of the corresponding ALE functions as $\mathbf{X}$ varies over the training data. For example, the sample version of ${ALE}_{j, M}$ is 
\begin{equation}\label{eq: estimator of ALE main VIM}
{\widehat{ALE}}_{j,M}=\frac{1}{n}\sum_{i=1}^{n}\left[{\hat{f}}_{j,ALE}\left(x_{i,j}\right)\right]^2=\frac{1}{n}\sum_{k=1}^{K}{n_j(k)\left[{\hat{f}}_{j,ALE}\left(z_{k,j}\right)\right]^2}.
\end{equation}

\par
To assess whether third- and higher-order interactions are present, we suggest computing the $R^2$-like statistic (using sample variances and the sample version of $f_{ALE,2}(\textbf{x})$)
\begin{equation} \label{eq: R2 like statistic}
R_{ALE,2}^2\equiv 1-\frac{\mathrm{Var}(f\left(\mathbf{X}\right)-f_{ALE,2}\left(\mathbf{X}\right))}{\mathrm{Var}(f\left(\mathbf{X}\right))},
\end{equation}
where $f_{ALE,2}\left(\mathbf{X}\right)=\mathbb{E}[f\left(\mathbf{X}\right)]+\sum_{j=1}^{d} {f_{j,ALE}(X_j)}+\sum_{j=1}^{d}\sum_{l=j+1}^{d}{f_{\{j,l\},ALE}(X_j,X_l)}$ is the second-order ALE approximation of $f\left(\mathbf{X}\right)$. By the ALE decomposition property, $f\left(\mathbf{X}\right)-f_{ALE,2}\left(\mathbf{X}\right)$ is the sum of all third- and higher-order ALE interactions (see \cite{Apley2020} for definition of high-order ALE interactions and the decomposition property). If $R_{ALE,2}^2\approx 1$, third- and higher-order interactions are negligible, in which case ${ALE}_{j,2}$ provides a reasonable measure of the total importance of $X_j$. However, if $R_{ALE,2}^2\ll 1$, ${ALE}_{j,2}$ is an incomplete measure of the total importance of $X_j$. In the next section, we propose two new total effect VIMs that implicitly take into account interactions of all orders.

\section{Path ALE (PALE) Total Effect Variable Importance Measures}\label{Section5PALEVIMs}
\par
Similar to ${ALE}_{j,M}$ and ${ALE}_{j,2}$, we define the  total effect VIMs as the variance of certain functions of $X_j$ that are related to, but different from, ALE functions. The primary challenge is how to define appropriate functions, the main considerations for which we first illustrate with some transparent toy examples.

\subsection{Some Concepts and Considerations for PALE Total Effect VIMs}\label{Section5.1MotivationPALE}
\begin{example} \label{example: path motivation 1 additive no noise}
\par
Suppose $d=2$ and $f\left(\mathbf{x}\right)=x_1+x_2$, where $X_1$ and $X_2$ are independent $\mathrm{Uniform}[0,1]$ random variables. A training sample of size $n=30$ is plotted in the left panel of Fig.\ \ref{fig: path motivation 1 additive}. To compute the sample versions ${\hat{f}}_{1,ALE}\left(x\right)$ and ${\widehat{ALE}}_{1,M}$, we partition the sample range of $\left\{x_{i,1}:i=1,2,\ldots, 30\right\}$ into $K=5$ intervals $\{N_1\left(k\right)=\left(z_{k-1,1},z_{k, 1}\right]:k=1,2,\ldots,5\}$, each with $n_1\left(k\right)=6$ training observations. For any $k\in\left\{1,2,\ldots,5\right\}$ and for each observation $\mathbf{x}_i$ that falls into interval $N_1(k)$, we define the individual local effect of $X_1$ as the finite difference $f\left(z_{k,1},x_{i,2}\right)-f\left(z_{k-1,1},x_{i, 2}\right)=z_{k,1}-z_{k-1,1}$, which are plotted as short line segments in the left panel of Fig.\ \ref{fig: path motivation 1 additive}. We then average the $n_1\left(k\right)=6$ individual local effects in $N_1(k)$ for each $k\in\left\{1,2,\ldots, 5\right\}$ and accumulate these $K=5$ averaged local effects via \eqref{eq: estimator of ALE main uncentered}, which, after centering via \eqref{eq: estimator of ALE main centered}, produces ${\hat{f}}_{1,ALE}\left(x\right)$. The main effect VIM ${\widehat{ALE}}_{1,M}$ is then computed as the sample variance of $\{{\hat{f}}_{1,ALE}\left(x_{i,1}\right), i=1, 2, \ldots, 30\}$ via \eqref{eq: estimator of ALE main VIM}. 

\par
When computing ${\hat{f}}_{1,ALE}\left(x\right)$, we first averaged the $n_1\left(k\right)$ local effects within each interval $N_1\left(k\right)$, and then pieced them together end-to-end (the accumulation). The main idea behind our PALE total effect VIMs is to piece together the individual segments to form a collection of $n_1\left(k\right)$ functions, the variance of which (with respect to both $X_j$ and the functions being randomly drawn from this collection, and after centering so they are all zero at $x_j=c$ in \eqref{eq: QPALE function c to x} and \eqref{eq: SPALE function c to x} below) will be the total effect VIM of $X_j$. Here, each function, which we refer to as a PALE function, is the accumulation of $K$ local effects, with one local effect chosen within each interval $N_1\left(k\right)$. The primary consideration is how to determine which local effects are chosen and accumulated to form each PALE function. 

\par
Because $f\left(\mathbf{x}\right)$ contains no interaction in this example, the $n_1\left(k\right)$ local effects within any interval are the identical values $z_{k,1}-z_{k-1,1}$ (left plot of Fig.\ \ref{fig: path motivation 1 additive}). Consequently, it does not matter which $K$ local effects (one within each interval) we piece together to form each of the $n_1\left(k\right)$ PALE functions, and the $n_1\left(k\right)$ functions are all the same function $x_1-c$ of $x_1$, as depicted in the right plot of Fig.\ \ref{fig: path motivation 1 additive}. Here, we take the centering constant $c$ to be the right endpoint $z_{3,1}$ of the interval that contains $\bar{x}_1$, where $\bar{x}_1$ is the average of $x_1$ over the $n$ sample points. Moreover, aside from slightly different centering, these PALE functions are all the same as ${\hat{g}}_{1,ALE}\left(x_1\right)$, so their variance is exactly the same as ${\widehat{ALE}}_{1,M}$, i.e., these VIMs satisfy property (\ref{property: total>main}).

\par
In contrast, if there were an interaction (as in Example \ref{example: path motivation 2 interaction} below), it would matter how we pieced together the individual local effects to form the PALE functions. For $d=2$, a natural way to piece together the local effects is to order them within each interval according to their corresponding $\mathbf{x}_{i,\backslash j}$ ($=x_{i,2}$ in this example) values. That is, the first path pieces together the local effects corresponding to the smallest $x_{i,2}$ value within each interval, the second path pieces together the local effects corresponding to the second smallest $x_{i,2}$ value within each interval, and so on. The numbers on the line segments in the left plot of Fig.\ \ref{fig: path motivation 1 additive} show the ordering of the local effects according to $x_{i,2}$. Since there is no interaction in this example, this method of forming the paths would produce the same path ALE functions shown in the right plot of Fig.\ \ref{fig: path motivation 1 additive} and would again satisfy property (\ref{property: total>main}). The next example considers what happens when there is an interaction. 

\begin{figure}[!ht]
\centering
 \includegraphics[width=\textwidth]{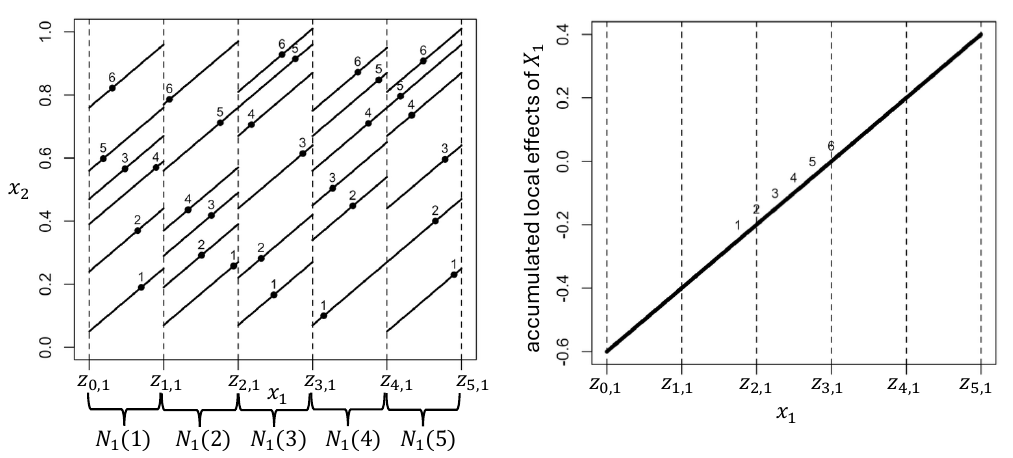}
    \caption{Illustration of the PALE functions for Example \ref{example: path motivation 1 additive no noise}, with no interaction. The bullets in the left plot are a scatter plot of $x_2$ vs. $x_1$ for the training data. The individual local effects are depicted as the line segments in the left plot, and the numbers on them represent their ordering in terms of $x_{i,2}$. The right plot shows the corresponding six PALE functions obtained by piecing together (accumulating) the local effects and centering, which are all identical because there is no interaction.}
    \label{fig: path motivation 1 additive}
\end{figure}

\end{example}

\begin{example}\label{example: path motivation 2 interaction}
\par
Suppose we again have $d=2$ with $X_1$ and $X_2$ following the same independent $\mathrm{Uniform}[0,1]$ distribution as in Example \ref{example: path motivation 1 additive no noise}, but with $f\left(\mathbf{x}\right)=x_1+x_2+2(x_1-0.5)\cdot(x_2-0.5)$ now including an interaction. The main-effect VIM ${\widehat{ALE}}_{1,M}$ can be calculated as in Example \ref{example: path motivation 1 additive no noise}. The $n_1\left(k\right)=6$ individual local effects within each interval $N_1\left(k\right)$ (plotted as short line segments in the left panel of Fig.\ \ref{fig: path motivation 2 interaction}) are now
\begin{equation}\label{eq: local effect of interaction toy example}
\begin{aligned}
f\left(z_{k,1},x_{i,2}\right)-f\left(z_{k-1,1}, x_{i,2}\right) 
&=\left[z_{k,1}+x_{i,2}+2\left(z_{k,1}-0.5\right)\cdot\left(x_{i,2}-0.5\right)\right]  \\
& -\left[z_{k-1,1}+x_{i,2}+2\left(z_{k-1,1}-0.5\right)\cdot\left(x_{i, 2}-0.5\right)\right]  \\
&=2(z_{k,1}-z_{k-1,1})\cdot x_{i,2},
\end{aligned}
\end{equation}
which is a function of $x_{i,2}$ because of the interaction. If we order the individual local effects within each interval according to $x_{i,2}$ (ordering is indicated by the numbers on the line segments in the left plot of Fig.\ \ref{fig: path motivation 2 interaction}) and accumulate them across the $K=5$ intervals, the resulting $6$ PALE functions are shown in the right plot of Fig.\ \ref{fig: path motivation 2 interaction}. We have again taken the centering constant $c$ to be the right endpoint $z_{3,1}$ of the interval containing $\bar{x}_1$. If we take the total effect VIM of $X_1$ to be the variance of the $6$ path ALE functions (with respect to $X_1$ and the functions being randomly drawn), it is now larger than the main effect VIM ${\widehat{ALE}}_{1,M}$, again satisfying property (\ref{property: total>main}). In Theorems 1 and 2 (Sections \ref{Section5.2QPALE} and \ref{Section5.3SPALE}), we show that the total effect VIM for $X_j$ computed as the variance of the collection of PALE functions always satisfies properties (\ref{property: total>main}) and (\ref{property: additive recovery}). 
\begin{figure}[!ht]
\centering
 \includegraphics[width=\textwidth]{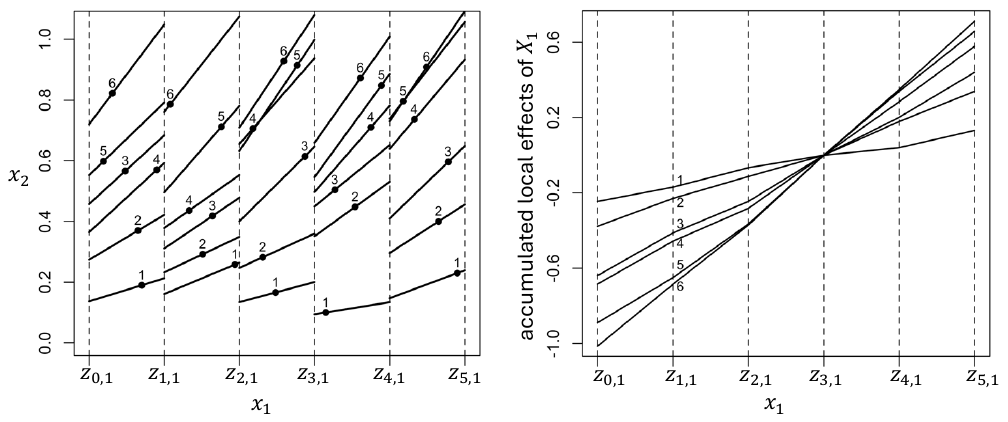}
    \caption{Illustration of the path ALE functions for Example \ref{example: path motivation 2 interaction}, which has an interaction. Compare to Fig.\ \ref{fig: path motivation 1 additive}, for which there is no interaction. The bullets in the left plot are a scatter plot of $x_2$ vs. $x_1$ for the training data. The six PALE functions in the right plot, which are obtained by accumulating the individual local effects in the left plot and then centering, are all different because of the interaction, and their variance is now larger than ${\widehat{ALE}}_{1,M}$.}
    \label{fig: path motivation 2 interaction}
\end{figure}
\end{example}

\par 
In the preceding examples with $d=2$, when forming the PALE functions to compute a total effect VIM for $X_1$, it seems reasonable to order the individual local effects within each interval according to the only other predictor, $X_2$. However, when $d>2$, there is no natural way to order the individual local effects. To illustrate some of the considerations involved, suppose $f\left(\mathbf{x}\right)=f_j\left(x_j\right)+f_{j,J}\left(x_j,\mathbf{x}_J\right)+f_{\backslash j}\left(\mathbf{x}_{\backslash j}\right)$, i.e.\ $X_j$ interacts with only a subset of predictors $\mathbf{X}_J\subseteq\mathbf{X}_{\backslash j}$. In this case, the local effect $f^j\left(x_j,\mathbf{x}_{\backslash j}\right)=\frac{df_j\left(x_j\right)}{dx_j}+\frac{\partial f_{j,J}\left(x_j, \mathbf{x}_J\right)}{\partial x_j}$ of $x_j$ is a function of $x_j$ and $\mathbf{x}_J$. Likewise, the sample version of the individual local effect associated with an observation $\mathbf{x}_i$ in interval $N_j(k)$ takes the form $f\left(z_{k,j}, \mathbf{x}_{i,\backslash j}\right)-f\left(z_{k-1,j},\mathbf{x}_{i,\backslash j}\right)=\left[f_j\left(z_{k,j}\right)+f_{j, J}\left(z_{k,j},\mathbf{x}_{i,J}\right)+f_{\backslash j}\left(\mathbf{x}_{i,\backslash j}\right)\right]-\left[f_j\left(z_{k-1,j}\right)+f_{j,J}\left(z_{k-1,j}, \mathbf{x}_{i,J}\right)+f_{\backslash j}\left(\mathbf{x}_{i,\backslash j}\right)\right]$ $= f_j\left(z_{k,j}\right)-f_j\left(z_{k-1,j}\right)+f_{j,J}\left(z_{k,j},\ \mathbf{x}_{i,J}\right)-f_{j,J}\left(z_{k-1,j},\mathbf{x}_{i,J}\right)$. Thus, the individual local effects within interval $N_j(k)$ vary due to the different values of $\mathbf{x}_{i,J}$ that fall into this interval. If we knew $J$ and $\left|J\right|=1$, we could order the individual local effects according to the values of the single variable $\mathbf{x}_{i,J}$. This cannot be done directly, however, since $J$ is unknown and we may have $\left|J\right|>1$. Our algorithm for finding the sample paths attempts to indirectly order the individual local effects according to $\mathbf{x}_{i,J}$.

\par
In Sections \ref{Section5.2QPALE} to \ref{Section5.4ConnectedSPALE}, we describe two approaches for ordering the individual local effects when forming the PALE functions that we have found to work well. We also more formally describe our total effect VIM framework. For notational simplicity, we present the results for differentiable $f(\cdot)$ with continuous numerical predictors having compact support. In Appendices \ref{Appendix: PALE definitions nondifferentiable f} and \ref{Appendix: handling categorical predictors} of the supplementary materials, we discuss how the VIMs are also applicable to non-differentiable $f(\cdot)$ and an extension to categorical predictors.

\subsection{Quantile Path ALE (QPALE) Total Effect VIM }\label{Section5.2QPALE}
\par
For predictor $X_j$ with support $\mathcal{S}_j=\left[x_{\min,j},x_{\max,j}\right]$, for each $z\in\mathcal{S}_j$ and $u\in\left[0,1\right]$, let $f_u^j\left(z\right)$ denote the $u$-quantile of the conditional distribution of $f^j\left(X_j,\mathbf{X}_{\backslash j}\right)|X_j=z$, i.e., such that $\mathbb{P}(f^j\left(X_j,\mathbf{X}_{\backslash j}\right)\le f_u^j\left(z\right)|X_j=z)=u$. Assume that $f_u^j\left(z\right)$ is Lebesgue integrable in $(z,u)$ on $\mathcal{S}_j\times\left[0,1\right]$ for the product Lebesgue measure. For each $\left(x, u\right)\in\mathcal{S}_j\times[0,1]$ and some constant $c\in\mathcal{S}_j$ chosen as described below, define the function 
\begin{equation}\label{eq: QPALE function c to x}
g_j\left(x,u;c\right)=\int_{c}^{x}{f_u^j\left(z\right)dz},
\end{equation}
which we term the quantile path accumulated local effects (QPALE) function of $X_j$. The convention in \eqref{eq: QPALE function c to x} is that for a function $h\left(z\right)$ and $x<c$, $\int_{c}^{x}h\left(z\right)dz=-\int_{x}^{c}h\left(z\right)dz$.

\par
The QPALE total effect VIM for $X_j$ is then defined as
\begin{equation} \label{eq: QPALE VIM definition}
{ALE}_{j,T}^Q=\mathrm{Var}\left[g_j\left(X_j,U;c\right)\right],
\end{equation}
where $U\sim\mathrm{Uniform}[0,1]$ is a uniform random variable that is independent of $X_j$. The constant $c$ can be chosen as $c = \mathbb{E}[X_j]$ or to minimize the variance of $g_j\left(X_j,U;c\right)$, i.e.\ $c= c^\ast=\argmin_c {\mathrm{Var}\left[g_j\left(X_j,U;c\right)\right]}$. We focus on the latter, although in our numerical experiments we have observed that $c^\ast \approx \mathbb{E}[X_j]$ typically.

\par
For each $u\in[0,1]$, the QPALE function $g_j\left(x,u;c\right)$ in \eqref{eq: QPALE function c to x} accumulates the $u$-quantile (with respect to the distribution of $\mathbf{X}_{\backslash j}$) of the local effect $f^j\left(X_j,\mathbf{X}_{\backslash j}\right)$ of $X_j$. In the context of Examples \ref{example: path motivation 1 additive no noise} and \ref{example: path motivation 2 interaction}, our sample version of ${ALE}_{j,T}^Q$ (described shortly) corresponds to ordering the individual local effects within each interval $N_j(k)$ according to the effects themselves and then accumulating these to form the PALE functions ${\hat{g}}_j\left(x,u;c\right)$. In Fig.\ \ref{fig: path motivation 2 interaction}, ordering according to $x_{i,2}$ did this implicitly. The following theorem, the proof of which is in Appendix \ref{Appendix: theorems and proofs} of the supplementary
materials, shows that ${ALE}_{j,T}^Q$ satisfies desirable properties (\ref{property: total>main}) and (\ref{property: additive recovery}).
\begin{thm1}
${ALE}_{j,T}^Q\geq{ALE}_{j,M}$ with equality if and only if $f(\cdot)$ is additive in $X_j$, i.e.\ $f\left(\mathbf{x}\right)=f_j\left(x_j\right)+f_{\backslash j}\left(\mathbf{x}_{\backslash j}\right)$, in which case ${ALE}_{j,T}^Q={ALE}_{j,M}=\mathrm{Var}\left[f_j\left(X_j\right)\right]$.  
\end{thm1}

\par
\textbf{Sample version of ${ALE}_{j,T}^Q$.} Computing the sample version of ${ALE}_{j,T}^Q$ is straightforward, since sample versions of the functions $g_j\left(x,u;c\right)$ are computed similarly to the functions ${\hat{g}}_{j, ALE}\left(x\right)$ in \eqref{eq: estimator of ALE main uncentered} that are used to compute the ALE main effect functions. Following the notation in Section \ref{Section4.1BackgroundALE}, we use the same partition $\{N_j\left(k\right)=\left(z_{k-1,j},z_{k, j}\right]:k=1,2,\ldots,K\}$ of the sample range of $\left\{x_{i,j}:i=1,2,\ldots, n\right\}$ with $N_j(k)$ containing $n_j(k)$ observations. Instead of averaging the individual local effects $f\left(z_{k,j}, \mathbf{x}_{i,\backslash j}\right)-f\left(z_{k-1,j},\mathbf{x}_{i,\backslash j}\right)$ as in the inner summation of \eqref{eq: estimator of ALE main uncentered}, we keep them separate and compute their quantiles, as follows. 

\par
Let $\Delta_j\left(k,u\right)$ denote the empirical $u$-quantile of the individual local effects in the $k$-th interval $N_j(k)$, i.e.\ $\Delta_j\left(k,u\right)$ is the $u$ sample quantile of $\{f\left(z_{k,j},\mathbf{x}_{i,\backslash j}\right)-f\left(z_{k-1,j},\mathbf{x}_{i,\backslash j}\right):i=1, 2, \ldots, n; x_{i,j}\in N_j(k)\}$. For each $x\in \left(z_{0,j}, z_{K,j}\right]$ and $u\in\left\{u_l=\left(l-1/2\right)/L: l=1,2, \ldots, L\right\}$, where $L$ is the number of quantile values to be considered, we use
\begin{equation}\label{eq: estimator of QPALE function from z0j to x}
{\hat{g}}_j\left(x,u;z_{0,j}\right)=\sum_{k=1}^{k_j\left(x\right)}{\Delta_j\left(k,u\right).}
\end{equation}
as the sample version of $g_j\left(x,u;z_{0,j}\right)$. Unless otherwise noted, we take the number of quantiles to be $L=n/K$, since there are $n_j(k)\approx n/K$ individual local effects within each interval. This results in computing $L=n/K$ QPALE functions, with each path associated with one of the quantiles. See Remark \ref{Remark: not one observation per leaf region}, later, and Appendix \ref{Appendix: handling categorical predictors} for how to handle the situation that $n_j(k)$ differs from $n/K$. 

\par
In Appendix \ref{Appendix: sample versions of PALE VIMs}, we provide a computationally efficient algorithm for finding the interval $N_j(k^\ast)$ in which the sample version of $c^\ast$ lies, with right endpoint $z_{k^\ast,j}$. We take the sample version, denoted by ${\widehat{ALE}}_{j,T}^Q$,  of ${ALE}_{j,T}^Q$ to be the sample variance of ${\hat{g}}_j\left(x_{i,j},u_l;z_{k^\ast,j}\right)={\hat{g}}_j\left(x_{i,j},u_l;z_{0,j}\right)-{\hat{g}}_j\left(z_{k^\ast,j},u_l;z_{0,j}\right)$ across $\{(x_{i,j}, u_l): i = 1, 2, \ldots, n; l = 1, 2, \ldots, L\}$, which can be efficiently computed via Eq.\ \eqref{eq: estimator QPALE VIM} in Appendix \ref{Appendix: sample versions of PALE VIMs}. The overall computational expense for computing ${\widehat{ALE}}_{j,T}^Q$ is modest and substantially less than for popular existing VIMs reviewed in Section \ref{Section3ReviewofVIMs}, which we illustrate in Section \ref{Section6.4ComputationalAdvantages}. 

\par
One drawback of ${\widehat{ALE}}_{j,T}^Q$ is that when $f(\cdot)$ is a noisy function (i.e., $f(\cdot)$ contains high-frequency random variation due to model training error), ${\widehat{ALE}}_{j,T}^Q$ can be artificially inflated due to the noise, which we illustrate with the following example and which motivates our second total effect VIM discussed in Section \ref{Section5.3SPALE}. 

\begin{example} \label{example: path motivation 3 additive with noise}
\par
In this modification of Example \ref{example: path motivation 1 additive no noise}, the two predictors $X_1$ and $X_2$ still follow an independent $\mathrm{Uniform}[0,1]$ distribution, and the training sample of size $n=30$ is plotted in the left panel of Fig.\ \ref{fig: path motivation 3 additive with noise with QPALE}. To illustrate how noise in $f(\cdot)$ can inflate ${\widehat{ALE}}_{j,T}^Q$, we now consider a noisy version $f\left(\mathbf{x}\right)=x_1+x_2+\mathrm{noise}$ of the underlying additive $\mathbb{E}\left[Y\middle|\mathbf{X}=\mathbf{x}\right]=x_1+x_2$, where the added noise is i.i.d. Using the same partition of $\left\{x_{i, 1}:i=1, 2, \ldots, 30\right\}$ into $K=5$ intervals, we calculate the individual local effects of $X_1$ within each interval via $f\left(z_{k,1},x_{i,2}\right)-f\left(z_{k-1,1}, x_{i,2}\right)=\left(z_{k,1}+x_{i, 2}+ \mathrm{noise}\right)-\left(z_{k-1,1}+ x_{i, 2}+\mathrm{noise}\right)=z_{k,1}-z_{k-1, 1}+\mathrm{noise}$. Comparing the left panels of Figs \ref{fig: path motivation 1 additive} and \ref{fig: path motivation 3 additive with noise with QPALE}, the local effects (represented by line segments) in Fig.\ \ref{fig: path motivation 3 additive with noise with QPALE} are also noisy and are no longer all the same value $z_{k,1}-z_{k-1,1}$ within each interval. If we form quantile paths according to the ordering of local effects within intervals, the resulting six QPALE functions are all substantially different functions, as shown in the right plot of Fig.\ \ref{fig: path motivation 3 additive with noise with QPALE} (using the same centering convention as in Figs \ref{fig: path motivation 1 additive} and \ref{fig: path motivation 2 interaction}), resulting in ${\widehat{ALE}}_{1,T}^Q\gg{\widehat{ALE}}_{1, M}$. This occurs because the path labeled 1 is
comprised of the smallest individual local effects within each interval, path 2 is comprised of the second
smallest, and so on. Technically, $f(\cdot)$ does contain an interaction, but it is only due to random noise, and the resulting inflation of ${\widehat{ALE}}_{1,T}^Q$ seems undesirable. Models like neural networks typically result in a relatively smooth $f(\cdot)$ if tuned properly, even when the underlying data to which $f(\cdot)$ is fit are noisy. For these models, we have not observed any substantial inflation in ${\widehat{ALE}}_{j,T}^Q$ in any of our examples. However, for non-smooth models such as random forests or boosted trees, the noise in $f(\cdot)$ sometimes does inflate ${\widehat{ALE}}_{j,T}^Q$. 

\begin{figure}[!ht]
\centering
 \includegraphics[width=\textwidth]{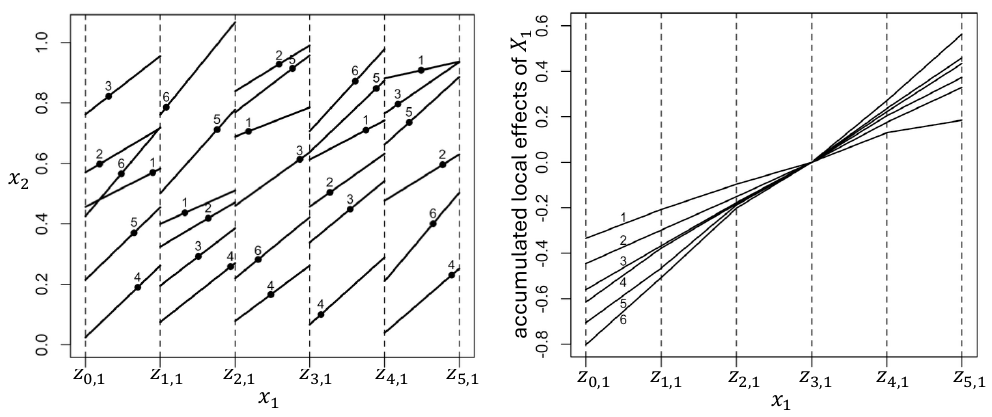}
    \caption{Illustration of how noise in $f(\cdot)$ inflates ${\widehat{ALE}}_{1,T}^Q$. Compared with Fig.\ \ref{fig: path motivation 1 additive} in which the individual local effects are all the same within each interval, the noise causes them to vary randomly (left plot). This results in the QPALE functions  being substantially different (right plot), which causes ${\widehat{ALE}}_{1,T}^Q\gg{\widehat{ALE}}_{1,M}$.}
    \label{fig: path motivation 3 additive with noise with QPALE}
\end{figure}

\par
If we instead form paths according to the ordering of $x_{i,2}$ values in each interval (see the left panel of Fig.\ \ref{fig: path motivation 4 additive with noise with X_backslash j}), the individual local effects that are accumulated to form each path function are more connected in the $\mathbf{X}_{\backslash j}$ (=$X_2$ in this example) space in the sense that the individual local effects in neighboring intervals along the same path will have more similar $x_{i,2}$ values. The right panel of Fig.\ \ref{fig: path motivation 4 additive with noise with X_backslash j} shows the accumulated individual local effects along these paths, which tend to smooth out the noise in the individual local effects relative to the quantile-based paths (compare the right plots of Figs.\ \ref{fig: path motivation 3 additive with noise with QPALE} and \ref{fig: path motivation 4 additive with noise with X_backslash j}). As a result, the variance of these path ALE functions will be smaller than ${\widehat{ALE}}_{1,T}^Q$ and closer to ${\widehat{ALE}}_{1,M}$, which seems more desirable when any interactions in $f(\cdot)$ are purely due to noise. 

\begin{figure}[!ht]
\centering
 \includegraphics[width=\textwidth]{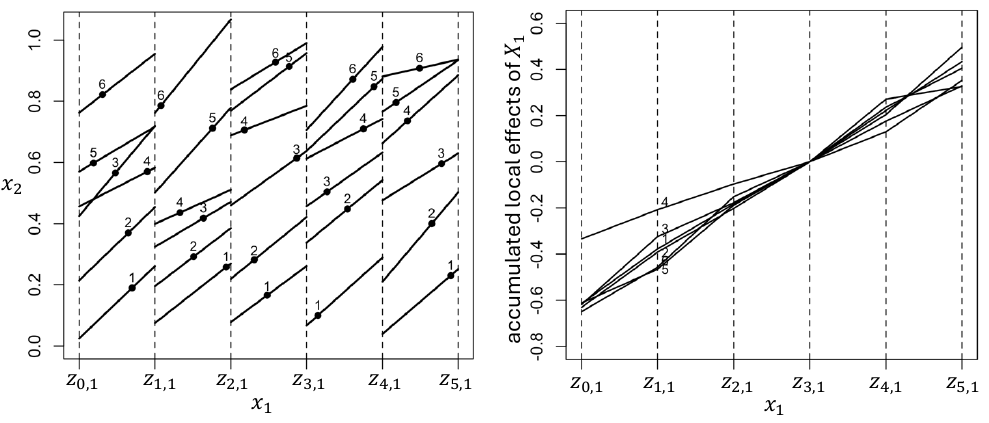}
    \caption{Illustration of how ordering the individual local effects according to their $\mathbf{x}_{i,\backslash j}$ values, as opposed to the quantile-based ordering, reduces the VIM inflation that occurs in ${\widehat{ALE}}_{1,T}^Q$ when $f(\cdot)$ is noisy. The PALE functions in the right plot have smaller variance than the QPALE functions in the right plot of Fig.\ \ref{fig: path motivation 3 additive with noise with QPALE}.}
    \label{fig: path motivation 4 additive with noise with X_backslash j}
\end{figure}

\end{example}

\subsection{Stochastic Path ALE (SPALE) Total Effect VIM}\label{Section5.3SPALE}
\par
Ordering the individual local effects according to their proximity in the $\mathbf{X}_{\backslash j}$ space when forming the PALE functions is trivial to do when $d=2$, since $\mathbf{X}_{\backslash j}$ is one-dimensional, and the individual local effects can be ordered according to their scalar $\mathbf{x}_{i,\backslash j}$ values. In Section \ref{Section5.4ConnectedSPALE} we propose a tree-based algorithm for ordering the individual local effects for any $d$ that encourages neighboring segments along each path to be more connected (i.e., in closer proximity) in the $\mathbf{X}_{\backslash j}$ space, and especially in the $\mathbf{X}_J$ space, where $\mathbf{X}_J$ denotes the unknown subset of predictors with which $X_j$ interacts.

\par
Before describing our path finding algorithm, we define a more general notion of stochastic paths that can be used in PALE total effect VIMs. These VIMs can be viewed as an abstraction of the QPALE total effect VIM and the VIM based on ordering the individual local effects according to $\mathbf{X}_{\backslash j}$ as in Examples \ref{example: path motivation 1 additive no noise} to \ref{example: path motivation 3 additive with noise}. To motivate the concepts, consider that the manner of ordering (based on either the quantiles or on $\mathbf{X}_{\backslash j}$) can be viewed as defining paths in $\mathbb{R}^d$ whereby, as $x_{i,j}$ varies over $\mathcal{S}_j$, the individual local effects $f\left(z_{k_j\left(x_{i,j}\right),j}, \mathbf{x}_{i,\backslash j}\right)-f\left(z_{k_j\left(x_{i,j}\right)-1,j}, \mathbf{x}_{i,\backslash j}\right)$ at position $(x_{i,j},\mathbf{x}_{i,\backslash j})$ along the paths have the following property: Across the collection of all paths, the $\mathbf{x}_{i,\backslash j}$ values that appear in the individual local effects are implicitly drawn from their conditional distribution $p_{\backslash j|j}\left(\cdot|x_{i,j}\right)$, since they are drawn from the data.

\par
To generalize this, we present the concepts in the context of a differentiable $f(\cdot)$ and describe in Appendix \ref{Appendix: PALE definitions nondifferentiable f} of the supplementary materials how the approach is also applicable to nondifferentiable $f(\cdot)$. Consider a differentiable supervised learning model $f(\cdot)$ and a predictor $X_j$ with support $\mathcal{S}_j=\left[x_{\min,j},x_{\max,j}\right]$. Let $\left(\Omega,\mathcal{F},P\right)$ denote some probability space with outcomes $\omega\in\Omega$ and measure $P$ (the details of which are unimportant, since we do not use it directly). For all $x\in\mathcal{S}_j$, define a stochastic path function $\mathbf{L}_x\left(\omega\right):\mathcal{S}_j\times\Omega\longmapsto\mathbb{R}^{d-1}$ as a stochastic process over domain $\mathcal{S}_j$ that satisfies the following properties:
\begin{itemize}
    \item[\textit{P1}] 
    For each $x\in\mathcal{S}_j$, $\mathbf{L}_x\left(\omega\right):\Omega\longmapsto\mathbb{R}^{d-1}$ is a $\left(d-1\right)-$variate random vector on $\left(\Omega,\mathcal{F},P\right)$ whose distribution is $p_{\backslash j|j}\left(\cdot|x\right)$. 
    
    \item[\textit{P2}] $f^j\left(x,\mathbf{L}_x\left(\omega\right)\right)$ is Lebesgue integrable in $x$ over $\mathcal{S}_j$ for almost all $\omega\in\Omega$ (i.e., for some set $A\in\mathcal{F}$ with $P\left(A\right)=1)$.
\end{itemize}
In other words, each path $\left\{\mathbf{L}_x\left(\omega\right):x\in\mathcal{S}_j\right\}$ can be viewed as a realization (determined by $\omega$) of a $\left(d-1\right)-$variate stochastic process on $\mathcal{S}_j$ whose marginal distribution at each $x$ is $p_{\backslash j|j}\left(\cdot|x\right)$, the conditional distribution of $\mathbf{X}_{\backslash j}|X_j=x$. In the following, the collection of paths $\left\{\left\{\mathbf{L}_x\left(\omega\right):x\in\mathcal{S}_j\right\}: \omega\in\Omega\right\}$ are the paths along which we will accumulate the local effects of $X_j$, where $\left\{\mathbf{L}_x\left(\omega\right):x\in\mathcal{S}_j\right\}$ represents the single path associated with that $\omega$.

\par
Given a stochastic path function $\mathbf{L}_\cdot\left(\cdot\right)$, for each $\left(x,\omega\right)\in\mathcal{S}_j\times\Omega$ and any constant $c\in\mathcal{S}_j$, define the function 
\begin{equation}\label{eq: SPALE function c to x} 
g_{j,\mathbf{L}}\left(x,\omega;c\right)=\int_{c}^{x}{f^j\left(z,\mathbf{L}_z\left(\omega\right)\right)dz},
\end{equation}
which we refer to as the stochastic path accumulated local effects (SPALE) function of $X_j$. The SPALE total effect VIM for $X_j$ is then defined as 
\begin{equation}\label{eq: SPALE VIM definition}
{ALE}_{j,T}^S=\mathrm{Var}\left[g_{j,\mathbf{L}}\left(X_j,\omega;c\right)\right],
\end{equation}
where the distribution of $\left(X_j,\omega\right)$ is taken to be their product measure. Similar to what we did in \eqref{eq: QPALE VIM definition}, we choose the constant $c$ to be $c^\ast=\argmin_c{\mathrm{Var}\left[g_{j,\mathbf{L}}\left(X_j,\omega;c\right)\right]}$ ($c = \mathbb{E}[X_j]$ could also be used). 

\par
Analogous to Theorem 1, the following theorem (proved in Appendix \ref{Appendix: theorems and proofs} of the supplementary
materials) shows that ${ALE}_{j,T}^S$ satisfies desirable properties (\ref{property: total>main}) and (\ref{property: additive recovery}).
\begin{thm2}
${ALE}_{j,T}^S\geq{ALE}_{j,M}$ with equality if and only if $f(\cdot)$ is additive in $X_j$, i.e.\ $f\left(\mathbf{x}\right)=f_j\left(x_j\right)+f_{\backslash j}\left(\mathbf{x}_{\backslash j}\right)$, in which case ${ALE}_{j,T}^S={ALE}_{j,M}=\mathrm{Var}\left[f_j\left(X_j\right)\right]$.  
\end{thm2}

\begin{remark} \label{Remark: measure of interaction with constant path}
Although the Theorem 2 statement that ${ALE}_{j,T}^S\geq{ALE}_{j,M}$ when $X_j$ interacts with other predictors is desirable, it would also be desirable if the difference between ${ALE}_{j,T}^S$ and ${ALE}_{j,M}$ increases as the strength of the interaction increases. This follows trivially from Eq.\ \eqref{eq: proof thm2 VIM_S ge main} in Appendix \ref{Appendix: theorems and proofs}, which shows that ${ALE}_{j,T}^S = {ALE}_{j,M} + \mathbb{E}\left[\mathrm{Var}\left[g_{j,\mathbf{L}}\left(X_j,\omega;c\right)|X_j\right]\right]$, if we define the second term as a measure of the strength of the $X_j$ interactions. This is an imperfect measure, since one could perhaps engineer paths $\mathbf{L}$ for which $\mathbb{E}\left[\mathrm{Var}\left[g_{j,\mathbf{L}}\left(X_j,\omega;c\right)|X_j\right]\right]$ does not necessarily increase as other measures of interaction strength increase (and vice versa). However, for appropriate choice of paths we view $\mathbb{E}\left[\mathrm{Var}\left[g_{j,\mathbf{L}}\left(X_j,\omega;c\right)|X_j\right]\right]$ as a reasonable measure of interaction strength based on Remark \ref{Remark: constant path continued} in Appendix \ref{Appendix: theorems and proofs}. 
\end{remark}

\par
\textbf{Sample version of ${ALE}_{j,T}^S$}. To compute the sample version of ${ALE}_{j,T}^S$, we first need the sample versions of the stochastic paths $\left\{\mathbf{L}_x\left(\omega\right):x\in\mathcal{S}_j; \omega\in\Omega\right\}$, where each path corresponds to different $\omega$. The only role of the sample stochastic paths is to dictate how we piece together the individual local effects to form the SPALE functions, analogous to how they were pieced together in Examples \ref{example: path motivation 1 additive no noise} to \ref{example: path motivation 3 additive with noise} by ordering them via either their quantiles or their $\mathbf{x}_{i,\backslash j}$ values. We refer to finding this ordering as finding the sample stochastic path function $\hat{\mathbf{L}}$ and defer its discussion until Section \ref{Section5.4ConnectedSPALE}. For now, assume we have $\hat{\mathbf{L}}$ available, which consists of a discrete collection of $L$ paths that we denote by $\left\{{\hat{\mathbf{L}}}\left(\omega_l\right): l=1,2,\ldots,L\right\}$, where each path $\hat{\mathbf{L}}(\omega_l) = \left\{\hat{\mathbf{L}}_k(\omega_l): k = 1, 2, \ldots, K\right\}$ consists of $K$ points in $\mathbb{R}^{d-1}$ with one point associated with each of the $K$ intervals that partition the sample range of $\left\{x_{i, j}:i=1, 2, \ldots, n\right\}$. Unless otherwise noted, as in the QPALE method, we take the number of paths to be $L=n/K$.

\par
Computing the sample version ${\widehat{ALE}}_{j,T}^S$ follows the same logic as for ${\widehat{ALE}}_{j,T}^Q$. Let $\Delta_{j,\mathbf{L}}\left(k,\omega_l\right)$ \\$=f\left(z_{k,j},{\hat{\mathbf{L}}}_k\left(\omega_l\right)\right)-f\left(z_{k-1,j},{\hat{\mathbf{L}}}_k\left(\omega_l\right)\right)$ denote the individual local effect of $X_j$ evaluated on path $\hat{\mathbf{L}}\left(\omega_l\right)$ in interval $N_j(k)$. The $\Delta_{j,\mathbf{L}}\left(k,\omega_l\right)$ quantities assume the same role as the individual local effects $\Delta_j\left(k,u_l\right)$ that are accumulated along the quantile paths in the QPALE method. For each $x\in \left(z_{0,j}, z_{K,j}\right]$ and for each $l=1, 2,\ldots, L$, we use
\begin{equation}\label{eq: estimator of SPALE function from z0j to x}
{\hat{g}}_{j,L}\left(x,\omega_l;z_{0,j}\right)=\sum_{k=1}^{k_j\left(x\right)}{\Delta_{j,\mathbf{L}}\left(k,\omega_l\right)}
\end{equation}
as the sample version of $g_{j,\mathbf{L}}\left(x,\omega_l;z_{0,j}\right)$.

After finding the interval $N_j(k^\ast)$ in which the sample version of $c^\ast$ lies (see Appendix \ref{Appendix: sample versions of PALE VIMs}) with endpoint $z_{k^\ast,j}$,  we take $\widehat{ALE}_{j,T}^S$ to be the sample variance of ${\hat{g}}_{j,\mathbf{L}}\left(x_{i,j},\omega_l;z_{k^\ast,j}\right)={\hat{g}}_{j,\mathbf{L}}\left(x_{i,j},\omega_l;z_{0,j}\right)-{\hat{g}}_{j,\mathbf{L}}\left(z_{k^\ast,j},\omega_l;z_{0,j}\right)$ across $\left\{(x_{i,j}, \omega_l ): i = 1, 2, \ldots, n; l = 1, 2, \ldots, L\right\}$ via Eq.\ \eqref{eq: estimator SPALE VIM} in Appendix \ref{Appendix: sample versions of PALE VIMs}.

\subsection{Connected Choice of SPALE Paths}\label{Section5.4ConnectedSPALE}
\par
Rather than attempt to characterize specific choices for the theoretical path $\mathbf{L}_x\left(\omega\right)$, we focus on choosing reasonable sample stochastic path functions $\hat{\mathbf{L}}$ to be used in ${\widehat{ALE}}_{j,T}^S$ that are in keeping with the spirit of Properties P1 and P2 and Remark \ref{Remark: measure of interaction with constant path}. The Lebesgue integrability condition in P2 loses some relevancy for $\hat{\mathbf{L}}$, because the finite differences $\Delta_{j,\mathbf{L}}\left(k,\omega_l\right)=f\left(z_{k,j},{\hat{\mathbf{L}}}_k\left(\omega_l\right)\right)-f\left(z_{k-1,j},\ {\hat{\mathbf{L}}}_k\left(\omega_l\right)\right)$ along the discretized sample paths (discretized into $K$ segments) are always Lebesgue integrable.  

\par
Assuming $n_j\left(k\right)=n/K$ is the same for each $k$, we use $L=n_j\left(k\right)$, and for each $k$, we take $\left\{{\hat{\mathbf{L}}}_k\left(\omega_l\right):l=1,2,\ldots,L\right\}$ to be the $L$ sample observations $\mathbf{x}_{i,\backslash j}$ that fall into interval $N_j(k)$ (i.e., for which $x_{i,j}\in N_j(k))$. This automatically satisfies the conditional distribution requirement P1, regardless of how the path segments are assigned to the $L$ paths. Our path finding procedure described in this section attempts to find a collection of sample paths $\hat{\mathbf{L}}$ that encourages the $\mathbf{x}_{i,\backslash j}$ values in neighboring segments ${\hat{\mathbf{L}}}_k\left(\omega_l\right)$ and ${\hat{\mathbf{L}}}_{k+1}\left(\omega_l\right)$ along the same path to be in proximity to each other, especially for the subset of predictors in $\mathbf{X}_{\backslash j}$ that truly interact with $X_j$. We refer to the resulting SPALE VIM as the connected PALE (CPALE) VIM and denote it by ${\widehat{ALE}}_{j,T}^C$. Such paths are more likely to (\textit{i}) result in ${\widehat{ALE}}_{j,T}^C-{\widehat{ALE}}_{j,M}$ increasing as the strength of the $X_j$ interactions increase, per Remark \ref{Remark: measure of interaction with constant path}; (\textit{ii}) mitigate the problem of noise in $f(\cdot)$ artificially inflating ${\widehat{ALE}}_{j,T}^C$ like it inflated ${\widehat{ALE}}_{j,T}^Q$ in Example \ref{example: path motivation 3 additive with noise}; and (\textit{iii}) satisfy the Lebesgue integrability condition in Property P2, in some hypothetical limit as $n\rightarrow\infty$ and $K\rightarrow\infty$ (which might be violated if the path segments were, say, randomly assigned to paths). 

\par
\begin{figure}[!ht]
\centering
 \includegraphics[width=\textwidth]{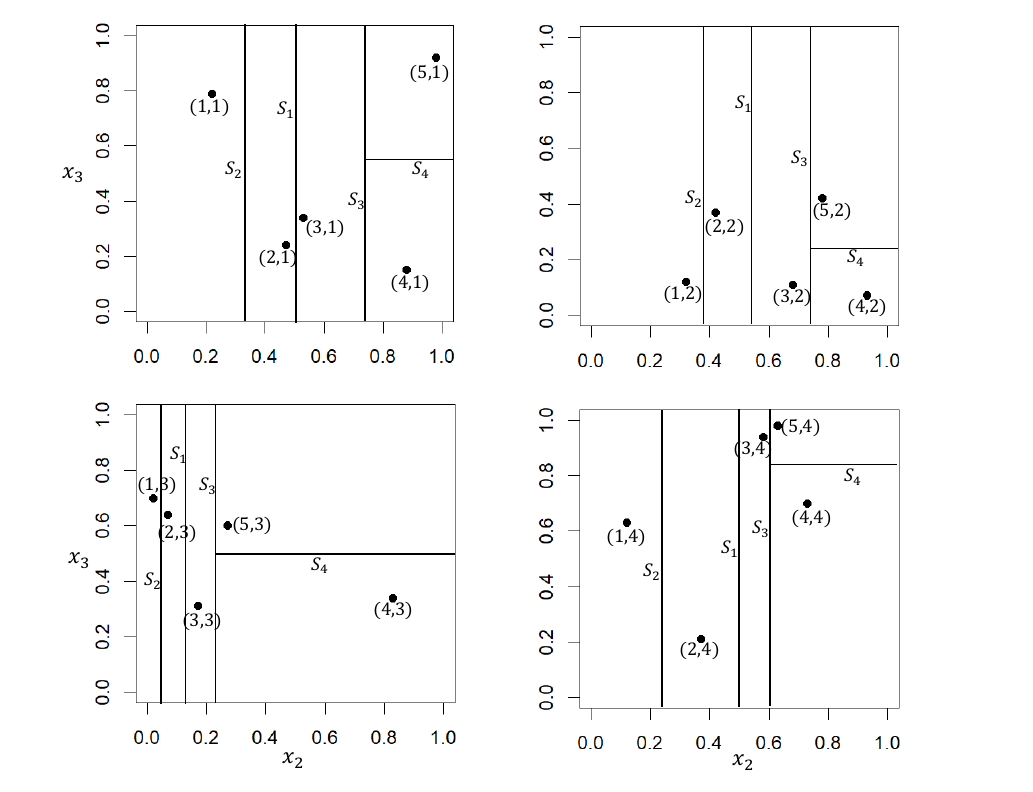}
    \caption{Illustration of the partitioning of the $\mathbf{x}_{i,\backslash j}$ sample space using Algorithm 1 recursively for an example with $d=3$, $n=20$, $X_j=X_1$, $K=4$ intervals, and $f\left(\mathbf{x}\right)=x_1+x_2+x_3+2x_1x_2$. The four plots are scatter plots of the $\mathbf{x}_{i,\backslash j} = (x_{i,2}, x_{i,3})$ values that fall into the $k$th interval, for $k=1,2,3,4$. The $4$ observations labeled $\left\{\left(l,k\right): k=1,2,3,4\right\}$ form the $l$th path, for $l=1, 2, 3, 4, 5$.}
    \label{fig: SPALE tree partitioning of x2-x3 space}
\end{figure}

For $d=2$, connected paths can be achieved simply by ordering the path segments according to their scalar $\mathbf{x}_{i,\backslash j}$ values, as we did in Examples \ref{example: path motivation 1 additive no noise} to \ref{example: path motivation 3 additive with noise}. Our path finding procedure attempts to do something analogous but for higher-dimensional $\mathbf{x}_{i,\backslash j}$. We use a form of tree partitioning of the $\mathbf{x}_{i,\backslash j}$ sample space within each interval, but we grow the tree simultaneously across all $K$ intervals in a manner that helps to ensure the $\mathbf{x}_{i,\backslash j}$ values associated with neighboring path segments along each path are in proximity to each other.  Fig.\ \ref{fig: SPALE tree partitioning of x2-x3 space} illustrates the tree partitioning procedure for an example with $d=3$, $n=20$, $X_j=X_1$, $K=4$ intervals, and $f\left(\mathbf{x}\right)=x_1+x_2+x_3+2x_1 x_2$. There are $n_j\left(k\right)=5$ observations within each interval, and so we must find $L=5$ paths.

\par
The following is an overview of the tree partitioning procedure for finding the paths $\hat{\mathbf{L}}$ to be used in ${\widehat{ALE}}_{j,T}^C$, and we provide pseudocode in Algorithm 1 with details for certain steps. For simplicity, we describe it for the situation for which $n_j\left(k\right)=n/K$ is the same for each $k$.
\begin{itemize}
    \item The procedure begins with each of the $K$ intervals consisting of a single region in the $\mathbf{X}_{\backslash j}$ space containing $n_j\left(k\right)$ observations, and all subsequent partitioning (i.e., splitting) occurs in the $\mathbf{X}_{\backslash j}$ space.
    \item 	For the first split, the $\mathbf{X}_{\backslash j}$ spaces in each of the $K$ intervals are simultaneously split on the same predictor (say $X_{m^\ast}$) chosen from among $\mathbf{X}_{\backslash j}$, and the locations of the $K$ simultaneous splits are always at the median of the $x_{i,m^\ast}$ values that fall into each interval. The predictor $X_{m^\ast}$ that is chosen for the split depends on the individual local effects in the left vs. right child branches of the split, the criterion for which is described in the last bullet below. In Fig.\ \ref{fig: SPALE tree partitioning of x2-x3 space}, the first such simultaneous split is labeled $S_1$ in each of the $K$ intervals, and the splits are on $X_{m^\ast}=X_2$. This first split produces a pair of child “leaf sets”, with the first leaf set comprised of the $K$ leaf regions in the left branch of $S_1$, and the second leaf set comprised of the $K$ leaf regions in the right branch. Algorithm 1 is pseudocode for splitting one existing leaf set into two child leaf sets, which is applied repeatedly (to the two children, and to the four grandchildren, and so on) in the steps below, where the $K$ leaf regions of the existing parent leaf set are denoted $(R_1,R_2, \ldots, R_K)$. Hypothetically, if there were only $n_j\left(k\right)=2$ observations in each interval, the path finding procedure would terminate here, with the first path being the $K$ $\mathbf{x}_{i,\backslash j}$ observations in the first leaf set and the second path being the $K$ $\mathbf{x}_{i,\backslash j}$ observations in the second leaf set.
    \item 	For each subsequent split, a single splitting variable is again chosen from among $\mathbf{X}_{\backslash j}$, and the $K$ leaf regions of one existing leaf set $(R_1,R_2,\ldots, R_K)$ are simultaneously split on this same predictor to produce two child leaf sets (a left branch and a right branch) using Algorithm 1, with the locations of the $K$ simultaneous splits always at the splitting variable’s median value among the $\mathbf{x}_{i,\backslash j}$ observations that fall into each leaf region. The same criterion for choosing the splitting variable, described in the last bullet, is used for every split. The subsequent splits of the two child leaf sets from the first split $S_1$ are labeled $S_2$ and $S_3$ in Fig.\ \ref{fig: SPALE tree partitioning of x2-x3 space}.  
    
    \item Algorithm 1 is applied repeatedly to continue splitting each existing leaf set $(R_1,R_2, \ldots, R_K)$ into two child leaf sets until there is only a single observation in each of the $K$ regions in each leaf set. This results in $L=n_j\left(k\right)$ leaf sets, and we take the $l$th path $(l=1,2,\ldots,L)$ to be $\left\{{\hat{\mathbf{L}}}_k\left(\omega_l\right)=\mathbf{x}_{i_{l,k},\backslash j}:k=1,2,\ldots,K\right\}$, where $i_{l,k}$ denotes the index of the single observation that falls into region $k$ of leaf set $l$. In Fig.\ \ref{fig: SPALE tree partitioning of x2-x3 space}, the four splits for this example are labeled $S_1$ to $S_4$, and the $K=4$ observations $\mathbf{x}_{i,\backslash j}$ comprising the $l$th path are labeled $\left\{\left(l,k\right):\ k=1,2,3,4\right\}$.
    \item 	When splitting each existing leaf set into two child leaf sets, we choose the splitting variable from among $\mathbf{X}_{\backslash j}$ as follows. For each $X_m$ of the $d-1$ potential splitting variables, we find the left and right child leaf sets as described above, by splitting on the sample medians of $X_m$ within each of the $K$ regions of the parent leaf set. The splitting variable $X_{m^\ast}$ is chosen as the one that maximizes $\sum_{k=1}^{K}\left|{\bar{\Delta}}_{m,k,left}-{\bar{\Delta}}_{m,k,right}\right|$, where ${\bar{\Delta}}_{m,k,left}$ (${\bar{\Delta}}_{m,k,right}$) denotes the average individual local effect of $X_j$ within region $k$ of the left (right) child leaf set. Choosing the splitting variable to maximize the difference between the average individual local effect in the left vs. right child regions tends to select splitting variables that have larger interactions with $X_j$. This helps to satisfy the connectivity goal by encouraging the $\mathbf{x}_{i,J}$ values in neighboring segments along the same path to be in proximity to each other, where $\mathbf{X}_J$ are the variables that interact most with $X_j$.
        
\end{itemize}

\begin{algorithm}[!ht]
\caption{Simultaneous partitioning of an existing leaf set $(R_1,R_2, \ldots, R_K)$ into two child leaf sets}\label{algorithm: CPALE tree partition step}
\begin{algorithmic}[1]
\Require{Parent leaf set $(R_1,R_2,\ldots, R_K)$:   $R_k\subseteq\left\{\mathbf{x}_{i,\backslash j}: x_{i,j}\in N_j\left(k\right),i=1, 2,\ldots,n\right\}$ is the leaf region in the $k$th interval 
at the current partitioning step.
} 
\For{each feature $X_m\in\mathbf{X}_{\backslash j}$}  
    \For{each $k=1, 2, \ldots,K$}
        \State{$M_{m,k}=$ median of $\{{x}_{i,m}:\mathbf{x}_{i,\backslash j}\in R_k\}$}
        \State{${\bar{\Delta}}_{m,k,\mathrm{left}}= $ average of $\{f\left(z_{k,j},\mathbf{x}_{i,\backslash j}\right)-f\left(z_{k-1,j},\mathbf{x}_{i,\backslash j}\right): \mathbf{x}_{i,\backslash j}\in R_k, x_{i,m}<M_{m,k}\}$ }
        \State{${\bar{\Delta}}_{m,k,\mathrm{right}}= $ average of $\{f\left(z_{k,j},\mathbf{x}_{i,\backslash j}\right)-f\left(z_{k-1,j},\mathbf{x}_{i,\backslash j}\right): \mathbf{x}_{i,\backslash j}\in R_k, x_{i,m}\geq M_{m,k}\} $ }

    \EndFor
\EndFor
\State{$m^\ast=\argmax_m\sum_{k=1}^{K}\left|{\bar{\Delta}}_{m,k,\mathrm{left}}-{\bar{\Delta}}_{m,k,\mathrm{right}}\right|; $ select any $m^\ast$ if not unique. }
\State{split $(R_1,R_2, \ldots, R_K)$ into child leaf sets $(R_{1,\mathrm{left}},R_{2,\mathrm{left}}, \ldots, R_{K,\mathrm{left}})$ and $(R_{1,\mathrm{right}},R_{2,\mathrm{right}}, \ldots, R_{K,\mathrm{right}})$, where the split of each $R_k$ $(k=1,2,\ldots,K)$ is on predictor $X_{m^\ast}$ at $M_{m^\ast,k}$ }
\Ensure{Two child leaf sets  $(R_{1,\mathrm{left}},R_{2,\mathrm{left}}, \ldots,R_{K,\mathrm{left}})$ and $(R_{1,\mathrm{right}},R_{2,\mathrm{right}}, \ldots, R_{K,\mathrm{right}})$}
\end{algorithmic}
\end{algorithm}

\begin{remark}\label{Remark: not one observation per leaf region}
There are situations in which it may not be desirable or possible to achieve a single observation within each final leaf set region. For example, if $n$ is not an integer multiple of $K$, we cannot set $n_j\left(k\right)=n/K$ to be the same value in each interval, in which case we cannot have a single observation in each final leaf set region. Additionally, if $n$ and $n_j\left(k\right)$ are extremely large, then we may prefer using $L<n_j\left(k\right)$. In either situation, the tree partitioning procedure could be terminated as soon as the number of paths reaches some specified value, and the procedure described in Appendix \ref{Appendix: handling categorical predictors} for handling varying $n_j(k)$ for categorical predictors (averaging the local effects within a leaf when $n_j(k) > L$ and reusing local effects on multiple paths when $n_j(k) < L$) can also be used here.
    
\end{remark}

\section{Examples and Discussion}\label{Section6Examples}

\subsection{Theoretical Comparison of VIMs for a Linear Model} \label{Section6.1TheoreticalExample}
\begin{example}
\label{example: theoretical VIMs of linear model}
\par
Suppose $d=3$, $\left(X_1, X_2\right)$ follows a bivariate standard normal distribution with correlation coefficient $\rho$, $X_3$ follows a standard normal distribution independent of $\left(X_1, X_2\right)$, and $f\left(\mathbf{X}\right)=\beta_1 X_1+\beta_2 X_2+\beta_3 X_3$. To remove the effects of the randomness of the sample versions and focus on the inherent nature of the VIMs, here we treat the true function $f\left(\mathbf{x}\right)$ and the distribution of $\mathbf{X}$ as known. In Appendix \ref{Appendix: derivations of example theoretical linear} we show that the various methods produce the following theoretical VIMs. The variance-based GSA total and main effect indices are ${GS}_{1,T}=\beta_1^2\left(1-\rho^2\right)$, ${GS}_{1,M}=\left(\beta_1+\rho\beta_2\right)^2$, ${GS}_{2,T}=\beta_2^2\left(1-\rho^2\right)$, ${GS}_{2,M}=\left(\rho\beta_1+\beta_2\right)^2$, and ${GS}_{3,T}={GS}_{3,M}=\beta_3^2$. If we take the global importance of $X_j$ to be $\mathbb{E}\left[\phi_j^2\left(\mathbf{X}\right)\right]$ and use $f_u\left(\mathbf{x}_u\right)=\mathbb{E}[f(\mathbf{x}_u,\mathbf{X}_{\backslash u})|\mathbf{X}_u=\mathbf{x}_u]$, the conditional global Shapley VIMs are ${SHC}_1=\left(\beta_1+\frac{\rho\beta_2}{2}\right)^2+\left(\frac{\rho\beta_1}{2}\right)^2-\rho^2\beta_1\left(\beta_1+\frac{\rho\beta_2}{2}\right)$, ${SHC}_2=\left(\beta_2+\frac{\rho\beta_1}{2}\right)^2+\left(\frac{\rho\beta_2}{2}\right)^2-\rho^2\beta_2\left(\beta_2+\frac{\rho\beta_1}{2}\right)$, and ${SHC}_3=\beta_3^2$. If we instead use $f_u\left(\mathbf{x}_u\right)=\mathbb{E}[f(\mathbf{x}_u,\mathbf{X}_{\backslash u})]$, the marginal global Shapley VIMs are ${SHM}_1=\beta_1^2$, ${SHM}_2=\beta_2^2$, and ${SHM}_3=\beta_3^2$. If we use the expected increase in squared error loss for the permutation VIMs, the conditional permutation VIMs are ${CP}_1=2\beta_1^2\left(1-\rho^2\right)$, ${CP}_2=2\beta_2^2\left(1-\rho^2\right)$, and ${CP}_3=2\beta_3^2$, and the marginal permutation VIMs are ${MP}_1=2\beta_1^2$, ${MP}_2=2\beta_2^2$, and ${MP}_3=2\beta_3^2$. The ALE-based VIMs for $X_j$ are ${ALE}_{j,T}^Q={ALE}_{j,T}^S={ALE}_{j,M}=\beta_j^2$. Fig.\ \ref{fig: theoretical VIMs of a linear model} plots the various total effect VIMs for $X_1$ vs. $\rho$ for $\beta_1=1$ and various $\beta_2$ ($\beta_3$ does not affect the VIMs for $X_1$ for any of the methods). 

\par
Notice that, regardless of $\rho$, desirable property (\ref{property: additive recovery}) is satisfied by the ALE VIMs since ${ALE}_{j,T}^Q={ALE}_{j,T}^S={ALE}_{j,M}=\beta_j^2$. The marginal global Shapley VIMs also produce ${SHM}_j=\beta_j^2$, as do the marginal permutation VIMs (aside from the factor of $2$), although they have the drawback of requiring extrapolation. Extrapolation was not an issue in this theoretical example because the true function $f\left(\mathbf{x}\right)$ was assumed known and can therefore reliably extrapolate. However, since $f\left(\mathbf{x}\right)$ must be estimated from training data in practice, extrapolation can be a serious drawback that degrades the marginal VIMs, especially when models more complex than linear models are fit. 

\par
When $\rho\neq 0$, none of the other methods produce VIMs equal to $\beta_j^2$. The conditional permutation VIMs are the same as the GSA total effect indices (aside from the factor of $2$), and the importances for both $X_1$ and $X_2$ decay to zero as $\rho$ approaches either $\pm1$. The conditional global Shapley VIMs also exhibit undesirable characteristics, in addition to having high computational expense. The importances (${SHC}_1$ and ${SHC}_2$) of the correlated predictors depend not only on the coefficient of the predictor but also on the coefficient of the other correlated predictor and on $\rho$. For example, when $\beta_2=0$ the predictor $X_2$ completely disappears from the model, but ${SHC}_2=\frac{\rho^2\beta_1^2}{4}$ is nonzero and is nearly as large as ${SHC}_1=\left(1-\frac{3\rho^2}{4}\right)\beta_1^2$ when $\left|\rho\right|$ is large.  Moreover, when $\beta_1=1$ and $\beta_3=0.5$ as in Figure \ref{fig: theoretical VIMs of a linear model}, for some combinations of $\rho$ and $\beta_2$, ${SHC}_1$ is smaller than ${SHC}_3=0.25$, even though $\beta_1$ is twice as large as $\beta_3$. Because we view these as fundamental problems with the GSA, conditional permutation, and conditional Shapley VIMs that make them inappropriate as VIMs, we omit them in the following numerical examples.

\begin{figure}[!ht]
\centering
 \includegraphics[trim={2cm 0cm 2cm 0cm},clip,width=\textwidth]{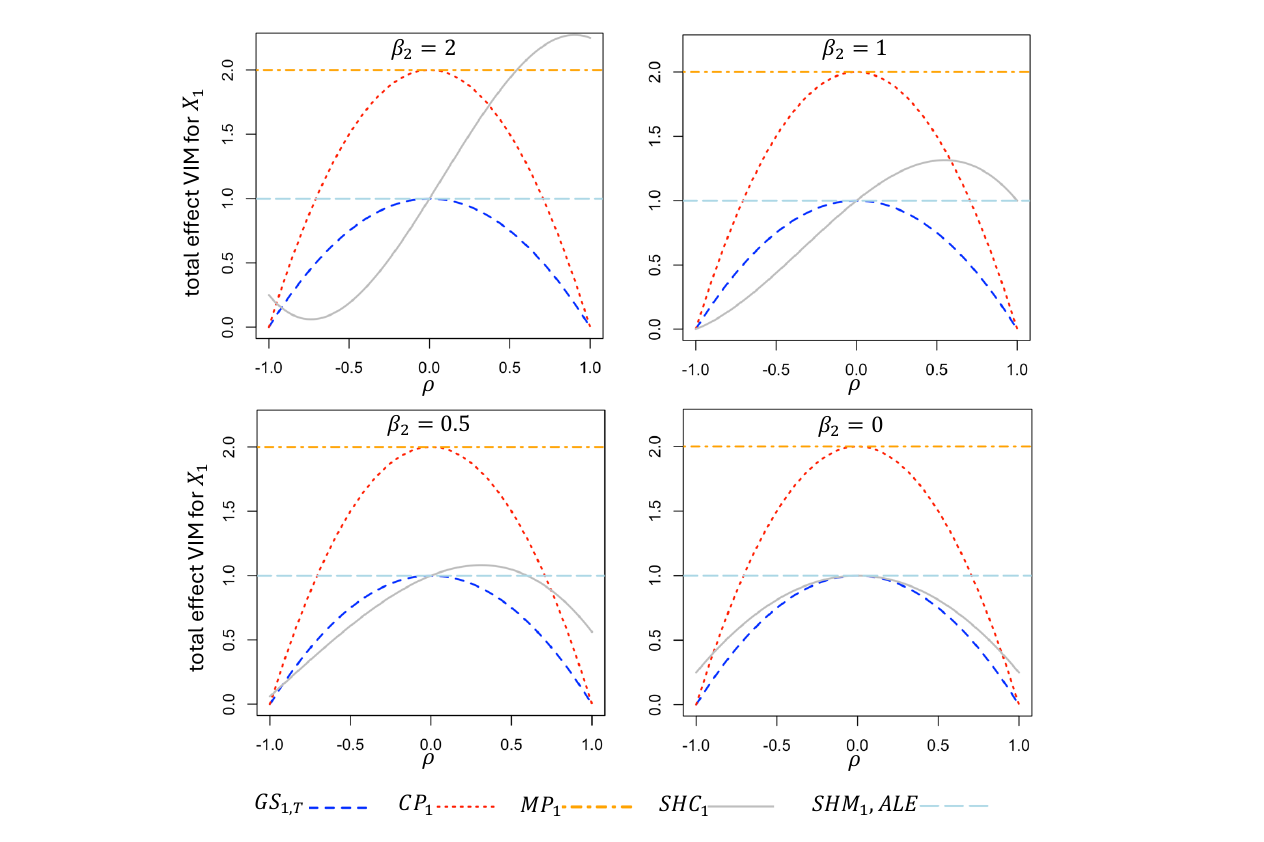}
    \caption{For the additive linear model of Example \ref{example: theoretical VIMs of linear model}, total effect VIMs for $X_1$ vs. $\rho$ for various $\beta_2$ with fixed $\beta_1=1$ and $\beta_3=0.5$.}
    \label{fig: theoretical VIMs of a linear model}
\end{figure}

\end{example}

\subsection{Numerical Comparisons of VIMs}\label{Section6.2NumericalExamples}
\par
In this subsection we present two numerical examples in which we generate training samples of predictors $\mathbf{X}$ from specified distributions $p(\mathbf{x})$, with the response generated via $Y=f_{\mathrm{true}}\left(\mathbf{x}\right)+\varepsilon$ for specified $f_{\mathrm{true}}\left(\mathbf{x}\right)$ and i.i.d.\ $\varepsilon\sim N(0, \sigma_\varepsilon^2)$ ($\sigma_\varepsilon=0.1$ in Example \ref{example: numerical example 2 predictors multiple replicates} and $\sigma_\varepsilon=0.5$ in Example \ref{example: numerical example 4 predictors interaction}). We fit neural network, gradient boosted tree, and random forest models to the simulated training samples using the \textbf{\texttt{nnet}} (\cite{Venables2002}), \textbf{\texttt{gbm}} (\cite{Ridgeway2015}), and \textbf{\texttt{randomForest}} (\cite{rForest}) packages of \texttt{R} and then apply the various VIM algorithms to the fitted models. In all examples, the model hyperparameters for each model were selected as approximately optimal according to multiple replicates of 10-fold cross validation to help ensure that the fitted models $f(\cdot)$ reasonably approximate $f_{\mathrm{true}}\left(\mathbf{x}\right)$ within the training data support.

\par
For each predictor $X_j$ and for each fitted model, we show the sample versions of the ALE main effect VIM (${ALE}_{j,M}$), PALE total effect VIM (${ALE}_{j,T}^Q$ and ${ALE}_{j,T}^C$), marginal permutation VIM measuring average increase in squared-error loss (${MP}_j$), and average squared marginal Shapley-values (${SHM}_j$) (corresponding to $f_u\left(\mathbf{x}_u\right)=\mathbb{E}[f(\mathbf{x}_u,\mathbf{X}_{\backslash u})]$), where the averages are across the $n$ training observations. The ${MP}_j$ and ${SHM}_j$ values were calculated using the \textbf{\texttt{iml}} (\cite{molnar2018iml}) package of \texttt{R}. For the random forest model, we also show the OOB marginal permutation importance (${MPO}_j$). Finally, for the two tree-based models, we also show the mean decrease in impurity ($MDI_j$, \cite{breiman1984cart}), which is a tree-based model-specific VIM that we did not review in Section \ref{Section3ReviewofVIMs}. For notational simplicity, we omit any circumflex symbols over all VIMs even though the reported values are the sample versions computed from the training data and fitted models $f(\cdot)$. In the following, the reported VIMs are always their square roots, so they are in the same units as $f$ and not $f^2$.  

\par
The following two examples illustrate how predictor correlation adversely affects the computed marginal VIMs (${MP}_j$, ${SHM}_j$, and ${MPO}_j$), even though their theoretical versions (analogous to those derived in Section \ref{Section6.1TheoreticalExample}) may be reasonable. In contrast, the ALE-based VIMs are less adversely affected by correlation and are more consistent with the theoretical results.

\begin{example}
\label{example: numerical example 2 predictors multiple replicates}
\par
There are $d=2$ predictors $(X_1,X_2)$ following a uniform distribution along the line segment $x_1=x_2$ from $0$ to $1$ with independent $N(0,{0.05}^2)$ added to both predictors. We generate $n=200$ observations of $\mathbf{X}$, and the true response function is $f_{\mathrm{true}}\left(\mathbf{x}\right)= x_1+x_2^2$. The error variance is $\sigma_\varepsilon^2={0.1}^2$, which results in theoretical $R^2=1-\frac{\mathrm{Var}\left(\varepsilon\right)}{\mathrm{Var}\left(Y\right)}\approx0.972$. We fit a neural network model to the training data and calculated VIMs for the fitted neural network. We repeated the experiment for a total of $100$ replicates, where on each replicate we generated a different training sample of size $n=200$, fitted a neural network to the training sample using the same (roughly optimal) tuning parameters for each replicate, and calculated the VIMs for the fitted model. There is no interaction in $f_{\mathrm{true}}$, and the additive terms are such that $\mathrm{Var}\left(X_1\right)\approx0.0858$ and $\mathrm{Var}\left(X_2^2\right)\approx0.0922$. As a result, desirable property (\ref{property: additive recovery}) implies that a reasonable ``ground truth" is to have the total effect VIM of $X_1$ about the same as (or slightly smaller than) the total effect VIM of $X_2$ and the main effect VIMs about the same as the total effect VIMs. 

\par
In Table \ref{table: numerical example 2 predictors multiple replicates}, we show the means and standard deviations of the VIMs across the $100$ Monte Carlo replicates. The mean VIMs for all methods follow the ``ground truth" patterns quite closely.  However, the replicate-to-replicate standard deviations for the ${MP}_j$ and ${SHM}_j$ VIMs for both predictors are much higher than for the ALE VIMs. The coefficients of variation (standard deviation divided by mean) for the ${MP}_j$ and ${SHM}_j$ VIMs are $2$ to $4$ times larger than for the ALE VIMs. The higher variability of the ${MP}_j$ and ${SHM}_j$ VIMs resulted in many replicates where the VIM for $X_1$ was much larger than for $X_2$, and vice versa, whereas the ALE VIMs for $X_1$ and $X_2$ were much more consistently similar across replicates. Table \ref{table: numerical example 2 predictors multiple replicates} also shows the absolute differences $|{VIM}_1-{VIM}_2|$ between the VIMs for $X_1$ versus $X_2$, with the differences being much larger for the ${MP}_j$ and ${SHM}_j$ VIMs than for the ALE VIMs. The $\left({VIM}_1,{VIM}_2\right)$ values for the replicate with the largest $|{VIM}_1-{VIM}_2|$ for each method are shown in the last row. 

\par
Altogether, the ALE-based VIMs agree much more closely (and more consistently across replicates) with the ``ground truth" than do the ${MP}_j$ and ${SHM}_j$ VIMs. The higher variability of the latter VIMs is likely the result of their need to extrapolate. As discussed in \cite{Apley2020}, the fitted neural network models have much higher prediction error when extrapolating outside the training data envelope than when predicting within the training data envelope, and the poor extrapolation accuracy has much less effect on the ALE VIMs.

\begin{table}[!ht]
\centering
\begin{tabular}{ |c | c c c c c| } 
 \hline
 & ${ALE}_{j,M}$ & ${ALE}_{j,T}^Q$ & ${ALE}_{j,T}^C$ & ${MP}_j$ & ${SHM}_j$\\ 
 \hline
 $X_1$ & 0.288 (0.036) & 0.290 (0.036) & 0.289 (0.034) & 0.474 (0.130) & 0.301 (0.067) \\ 
 $X_2$ & 0.307 (0.031) & 0.309 (0.030) & 0.308 (0.031) & 0.513 (0.187) & 0.328 (0.132) \\ 
 $\left|{VIM}_1-{VIM}_2\right|$ &  0.051 (0.043) & 0.051 (0.043) & 0.050 (0.042) & 0.139 (0.192) & 0.127 (0.134) \\
 most extreme & (0.202, 0.379) & (0.202, 0.379) & (0.202, 0.379) & (0.201, 0.930) & (0.689, 0.005) \\
 \hline
\end{tabular}
\caption{Square root of VIM results for the neural network model across $100$ replicates for Example \ref{example: numerical example 2 predictors multiple replicates} showing much better consistency across replicates for the ALE VIMs, relative to the ${MP}_j$ and ${SHM}_j$ VIMs that must extrapolate. The numbers in the first three rows are the means across all replicates with standard deviations in parentheses. The last row shows the $\left({VIM}_1,{VIM}_2\right)$ values for the replicate with the largest $|{VIM}_1-{VIM}_2|$ for each method.}
\label{table: numerical example 2 predictors multiple replicates}
\end{table}

\par
Even though the neural network models considered in this example all had cross validation $R^2$ similar to the theoretical $R^2$, indicating the models are accurate within the training data envelope, the relatively small sample size $n=200$ likely contributes to the inaccurate extrapolation. The next example uses much larger sample size $n$, in which case the ${MP}_j$ and ${SHM}_j$ have much smaller variability that is comparable to the ALE VIMs across replicates. We therefore report results for only one typical replicate. Nonetheless, other problems caused by extrapolation still remain and are discussed below. 
\end{example}

\begin{example}
\par 
For this example, we have $d=4$ predictors, and the joint distribution of $\mathbf{X}$ is modelled as a Gaussian copula (\cite{nelsen2006introcopula}) so that the marginal distribution of each predictor is $\mathrm{Uniform}\left[0,1\right]$ and $\rho_{13}=0.2$, $\rho_{23}=0.9$, $\rho_{12}=\rho_{14}=\rho_{24}=\rho_{34}=0$, where $\rho_{ij}$ is the correlation coefficient between $X_i$ and $X_j$. We used the \textbf{\texttt{copula}} package of \texttt{R} (\cite{hofert2024copula}) to generate the $n=10,000$ observations of $\mathbf{X}$. The response is generated with  $f_{\mathrm{true}}\left(\mathbf{x}\right)= 4x_1 + 3.87x_2^2 + 2.97\frac{\exp\left(-5+10x_3\right)}{1+\exp\left(-5+10x_3\right)}+ 13.86\left(x_1-0.5\right)\left(x_2-0.5\right)$. The error variance is $\sigma_\varepsilon^2={0.5}^2$, and the resulting theoretical $R^2=1-\mathrm{Var}(\varepsilon)/\mathrm{Var}\left(Y\right)\approx0.97$. The coefficients in $f_{\mathrm{true}}\left(\mathbf{x}\right)$ are chosen so that all additive terms have the same variance, i.e., $\mathrm{Var}\left(4X_1\right)=\mathrm{Var}\left(3.87X_2^2\right)=\mathrm{Var}\left(2.97\frac{\exp\left(-5+10X_3\right)}{1+\exp\left(-5+10X_3\right)}\right)=\frac{4}{3}$. The interaction term $13.86\left(X_1-0.5\right)\left(X_2-0.5\right)$ is designed so that its ALE main effect is $0$ and its variance is also $\frac{4}{3}$. As a result, we might consider the ``ground truth" to be that the main effect importances of $X_1$, $X_2$, and $X_3$ are all about the same and the total effect importance of $X_1$ and of $X_2$ are larger than that of $X_3$ due to the $\left(X_1,X_2\right)$ interaction. Since $X_4$ does not appear in  $f_{\mathrm{true}}\left(\mathbf{x}\right)$, its VIMs should be close to zero by desirable property (\ref{property: doesnotappear}). 

\par
Tables \ref{table: numerical example 4 predictors neural network results} to \ref{table: numerical example 4 predictors random forest results} show VIMs for each predictor $X_j$ calculated for the neural network, gradient boosted tree, and random forest models, respectively. For all three models, the ALE main effect and PALE total effect VIMs closely follow the ``ground truth" described above, except that the ${ALE}_{j,T}^Q$ values for the random forest and gradient boosted tree models are somewhat inflated. As discussed in Section \ref{Section5PALEVIMs}, the quantile paths tend to inflate the total effect importance for non-smooth models such as the two tree-based models here, which served as the motivation for the connected PALE VIM. Tables \ref{table: numerical example 4 predictors gradient boosted tree results} and \ref{table: numerical example 4 predictors random forest results} indicate that ${ALE}_{j,T}^C$ does correct this VIM inflation. 

\par
The $MP_j$ and $SHM_j$ VIMs for the neural network model in Table \ref{table: numerical example 4 predictors neural network results} agree with the PALE VIM results and follow the ``ground truth" pattern $VIM_1\approx VIM_2>VIM_3$, and $VIM_4\approx0$. However, the $MP_j$ and $SHM_j$ values for the two tree-based models in Tables \ref{table: numerical example 4 predictors gradient boosted tree results} and \ref{table: numerical example 4 predictors random forest results} do not follow this pattern. For both models, $MP_2$ and $SHM_2$ are somewhat deflated and $MP_3$ and $SHM_3$ are somewhat inflated compared to the corresponding values in Table \ref{table: numerical example 4 predictors neural network results}. In particular, they tend to indicate that $X_3$ is more important than $X_2$ and sometimes more important than $X_1$, and they do not assign the same importance to $X_1$ and $X_2$. For the random forest in Table \ref{table: numerical example 4 predictors random forest results}, the ${MPO}_j$ VIMs are also misleading (almost all the importance is assigned to $X_1$), as are the tree-specific $MDI_j$ VIMs in Tables \ref{table: numerical example 4 predictors gradient boosted tree results} and \ref{table: numerical example 4 predictors random forest results} ($X_3$ is assigned much more importance than $X_1$ and $X_2$). 

\begin{table}[!ht]
\centering
\begin{tabular}{ |c | c c c c c| } 
 \hline      
 & ${ALE}_{j,M}$ & ${ALE}_{j,T}^Q$ & ${ALE}_{j,T}^C$ & ${MP}_j$ & ${SHM}_j$\\ 
 \hline
 $X_1$ & 1.156 & 1.627 & 1.633 & 2.306 & 1.311 \\ 
 $X_2$ & 1.174 & 1.627 & 1.633 & 2.322 & 1.336 \\ 
 $X_3$ & 1.139 & 1.140 & 1.140 & 1.578 & 1.130  \\
 $X_4$ & 0.014 & 0.022 & 0.021 & 0.031 & 0.014 \\
 \hline
\end{tabular}
\caption{Square root of VIM results for the neural network model for Example \ref{example: numerical example 4 predictors interaction}.  }
\label{table: numerical example 4 predictors neural network results}
\end{table}

\begin{table}[!ht]
\centering
\begin{tabular}{ |c | c c c c c c| } 
 \hline
 & ${ALE}_{j,M}$ & ${ALE}_{j,T}^Q$ & ${ALE}_{j,T}^C$ & ${MP}_j$ & ${SHM}_j$ & $MDI_j$\\ 
 \hline
 $X_1$ & 1.176 & 1.786 & 1.652 & 2.293 & 1.301 & 25.03 \\ 
 $X_2$ & 1.150 & 1.635 & 1.564 & 2.043 & 1.200 & 13.01 \\ 
 $X_3$ & 1.155 & 1.322 & 1.169 & 1.767 & 1.230 & 61.71 \\
 $X_4$ & 0.023 & 0.139 & 0.036 & 0.099 & 0.035 & 0.26  \\
 \hline
\end{tabular}
\caption{Square root of VIM results for the gradient boosted tree model for Example \ref{example: numerical example 4 predictors interaction}. }
\label{table: numerical example 4 predictors gradient boosted tree results}
\end{table}

\begin{table}[!ht]
\centering
\begin{tabular}{ |c | c c c c c c c| } 
 \hline
 & ${ALE}_{j,M}$ & ${ALE}_{j,T}^Q$ & ${ALE}_{j,T}^C$ & ${MP}_j$ & ${SHM}_j$ & $MDI_j$ & ${MPO}_j$\\ 
 \hline

 $X_1$ & 1.208 & 2.171 & 1.685 & 2.292 & 1.276 & 21698.9 & 609.7 \\ 
 $X_2$ & 1.188 & 2.143 & 1.648 & 1.865 & 1.049 & 17288.4 & 58.3 \\ 
 $X_3$ & 1.218 & 1.773 & 1.237 & 2.007 & 1.370 & 40400.0 & 71.8 \\
 $X_4$ & 0.009 & 0.213 & 0.041 & 0.154 & 0.017 & 226.5 & 0.3 \\
 \hline
\end{tabular}
\caption{Square root of VIM results for the random forest model for Example \ref{example: numerical example 4 predictors interaction}.}
\label{table: numerical example 4 predictors random forest results}
\end{table}

\par
It is important to note that the ALE VIMs are quite consistent across the three models in Tables \ref{table: numerical example 4 predictors neural network results} to \ref{table: numerical example 4 predictors random forest results}, whereas the other VIMs give different results on the relative importance of $X_1$, $X_2$, and $X_3$, depending on which model one considers. This is a significant advantage of ALE VIMs, since the three models in this example are all equally good in terms of having similar cross validation $R^2\approx0.97$, which is almost identical to the theoretical $R^2$. Model accuracy within the training data support (which is what cross-validation assesses) is all that is needed for the ALE VIMs to capture the correct predictor importances, whereas ${MP}_j$ and ${SHM}_j$ also require model accuracy when extrapolating outside the data envelope. When we generated a test data set outside the data envelope by using independent predictor distributions, we found that the boosted tree and especially the random forest extrapolated much less accurately than the neural network in this example (results are omitted for brevity), which likely explains why ${MP}_j$ and ${SHM}_j$ for the boosted tree and random forest were less accurate than for the neural network.

\label{example: numerical example 4 predictors interaction}
\end{example}

\subsection{Bike Sharing Real Data Example}\label{Section6.3BikeSharing}
\begin{example}
\label{example: bikesharing example}
\par
 We now illustrate ALE VIMs for a bikeshare example, the data for which (\cite{BikeSharing}) can be found at the UCI machine learning repository and contain hourly bike sharing rental counts from the Capital Bikeshare System in Washington D.C, USA over 2011-2012, together with hourly weather and seasonal information for the same time period. After data cleaning, there are $n=17,333$ observations corresponding to $17,333$ hours and $d=10$ predictors: quarter ($X_1$, treated as numerical: $\{1, 2, \ldots, 8\}$), month ($X_2$, treated as numerical: 1 = January, 2 = February, $\ldots$, 12 = December), hour ($X_3$, treated as numerical: $\{0, 1, \ldots, 23\}$), holiday ($X_4$, categorical: 0 = non-holiday, 1 = holiday), weekday ($X_5$, treated as numerical: $\{0, 1, \ldots,6\}$ representing day of a week with 0 = Sunday), workingday ($X_6$, categorical: 0 = weekend or holiday, 1 = otherwise), weather situation ($X_7$, treated as numerical: $\{1,2,3,4\}$, with smaller values corresponding to better weather situations), atemp ($X_8$, numerical: feeling temperature in Celsius), hum ($X_9$, numerical: humidity), and windspeed ($X_{10}$, numerical: wind speed). The response is the log of the hourly bike rental counts. Some of the ten predictors are highly correlated, e.g. (month, atemp) and (weekday, holiday, workingday). We fit a neural network model with 35 nodes in the single hidden layer, a linear output activation function, and regularization parameter of 0.05 to predict the log hourly bike rental counts, with the model hyperparameters selected as approximately optimal via cross validation. The cross validation $R^2$ for the neural network is around $0.93$ (after converting the predicted response back to hourly rental counts instead of their logs), indicating that the model is reasonably accurate and appropriate to interpret. 
 
 \par
 In Table \ref{table: bikeshare results}, we show the ${ALE}_{j,M}$, ${ALE}_{j,T}^Q$, ${ALE}_{j,T}^C$, ${MP}_j$, and ${SHM}_j$ VIMs for all ten predictors for the neural network model. As before, we report the square roots of all VIMs so they are in the same units as $f$. The PALE, permutation, and Shapley VIMs all agree that hour ($X_3$) is by far the most important predictor for hourly rental counts, which makes sense intuitively. The ${ALE}_{j,T}^Q$, ${ALE}_{j,T}^C$ and ${MP}_j$ VIMs also agree that working day ($X_6$) is the second most important predictor, followed by the predictors atemp ($X_8$) and quarter ($X_1$),  although the relative orderings of atemp and quarter are slightly different. The ${SHM}_j$ VIM indicates that the same four predictors have the highest importance but that workingday is less important than quarter and atemp. All total effect VIMs are in agreement that the remaining six predictors are less important, with their relative orderings quite similar for PALE and permutation VIMS but somewhat different for Shapley VIMs.
 
\par    
       
\begin{table}[!ht]
\centering
\begin{tabular}{ |r | c c c c c| } 
 \hline
 & ${ALE}_{j,M}$ & ${ALE}_{j,T}^Q$ & ${ALE}_{j,T}^C$ & ${MP}_j$ & ${SHM}_j$\\ 
 \hline     
hour $X_3$ &1.233 & 1.982 & 1.983 & 1.870 & 1.269\\
workingday $X_6$ &0.090 & 0.559 & 0.561 & 0.671 & 0.240\\ 
atemp  $X_8$ &0.262 & 0.363 & 0.363 & 0.401 & 0.245\\
quarter $X_1$ &0.261 & 0.289 & 0.287  & 0.419 & 0.281 \\ 
holiday $X_4$  &0.025 & 0.265 & 0.268  & 0.287 & 0.058 \\
month $X_2$ &0.109 & 0.222 & 0.223 & 0.286 & 0.146\\ 
weekday $X_5$ &0.071 & 0.226 & 0.223  & 0.299 & 0.117 \\ 
hum $X_9$  &0.076 & 0.136 & 0.135  & 0.210 & 0.109\\ 
windspeed $X_{10}$ & 0.034 & 0.079  & 0.082  & 0.117 & 0.049 \\
weathersit $X_7$ &0.061 & 0.076 & 0.074 & 0.202 & 0.118 \\
 \hline
\end{tabular}
\caption{Square root of VIM results in descending order of ${ALE}_{j,T}^C$ for the neural network model for Example \ref{example: bikesharing example}.}
\label{table: bikeshare results}
\end{table}

\par
This example also illustrates an additional advantage of the ALE VIMs, which produce a main effect VIM that can be compared to the total effect VIM to gauge the importance of interactions. Table \ref{table: bikeshare results} shows that workingday $X_6$ has small main effect importance ${ALE}_{j,M}$ but the second largest total effect importance ${ALE}_{j,T}^C$, which indicates that it interacts strongly with some other predictors. To identify other predictor(s) with which $X_6$ interacts, we can construct second-order ALE interaction plots of $X_6$ paired with each of the other predictors and/or compute the sample variance of the second-order ALE interaction functions $f_{\{6,j\}, ALE}\left(x_{i6},x_{ij}\right)$ for $j\in\left\{1,2,\ldots,10\right\}\backslash \{6\}$. This revealed that the strongest $X_6$ interaction is with $X_3$ hour. Figure \ref{fig: bikeshare interaction plots} shows main effect ALE plots of $\bar{f}+\hat{f}_{j, ALE}\left(x_j\right)$ for $j=6, 3$, the pure second-order interaction effect ${\hat{f}}_{\{6,3\}, ALE}\left(x_6, x_3\right)$, and the ALE second-order approximation of the $\left(X_6,X_3\right)$ joint effect $\bar{f}+\hat{f}_{6,ALE}\left(x_6\right)+{\hat{f}}_{3,ALE}\left(x_3\right)+{\hat{f}}_{\{6,3\}, ALE}\left(x_6, x_3\right)$. The pure interaction plot indicates strong $\left(X_6,X_3\right)$ interaction, and the joint effect plot reveals the nature of the interaction and has clear interpretations. On working days the morning and evening rush hour peaks seen in the main effect ALE plot of $X_3$ are more pronounced, whereas on nonworking days they completely disappear.

\begin{figure}[!ht]
\centering
 \includegraphics[width=\textwidth]{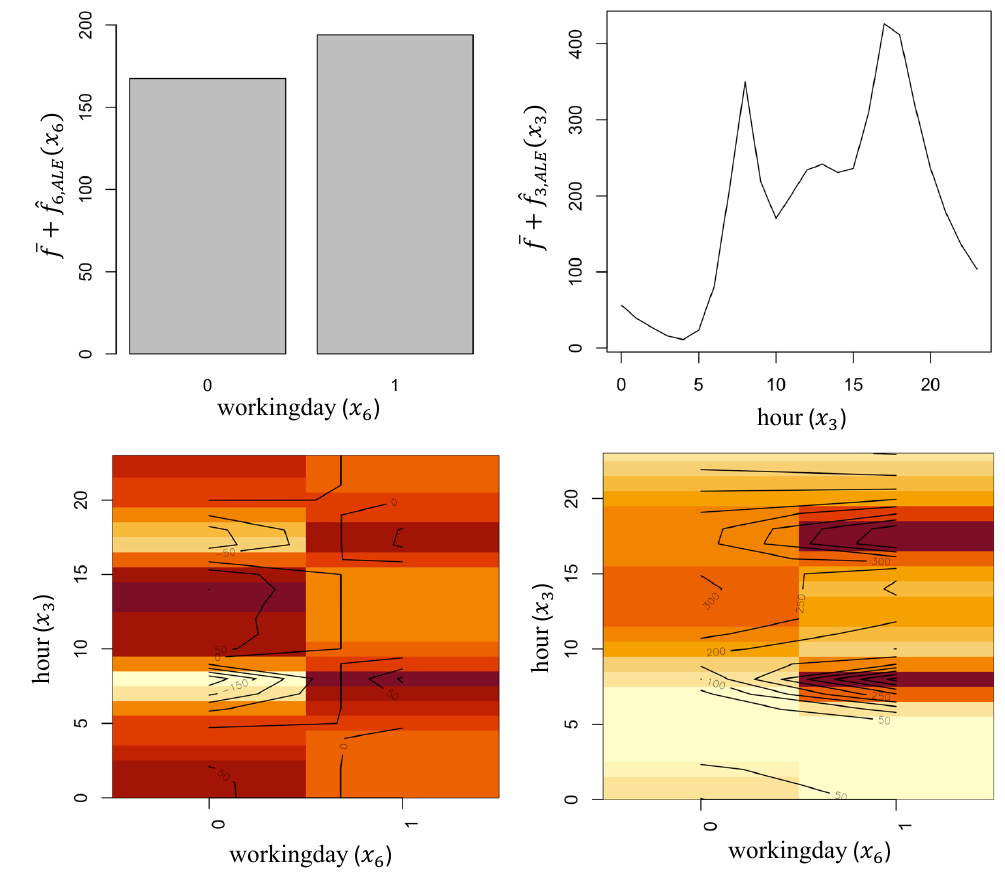}
    \caption{Main effect ALE plots of $\bar{f}+\hat{f}_{j, ALE}\left(x_j\right)$ for $j=6,3$ (top row), pure second-order interaction effect $\hat{f}_{\{6,3\}, ALE}\left(x_6, x_3\right)$ and ALE second order approximation of the $\left(X_6,X_3\right)$ joint effect $\bar{f}+\hat{f}_{6, ALE}\left(x_6\right)+\hat{f}_{3, ALE}\left(x_3\right)+\hat{f}_{\{6,3\}, ALE}\left(x_6, x_3\right)$ vs. $\left(x_6,x_3\right)$ (bottom row) for the bike sharing data example.}
    \label{fig: bikeshare interaction plots}
\end{figure}

\end{example}

\subsection{Computational Advantages}\label{Section6.4ComputationalAdvantages}
\par
We now analyze the time complexity of computing VIMs for each individual predictor $X_j$ for a supervised learning model $f(\cdot)$ fit to training data of size $n$ with $d$ predictors. As we discussed in Appendix \ref{Appendix: sample versions of PALE VIMs}, the computation of ${ALE}_{j,M}$, ${ALE}_{j,T}^Q$ and ${ALE}_{j,T}^C$ each requires $2n$ evaluations of $f(\cdot)$. Let $h(\cdot)$ denote the time complexity of one evaluation of $f(\cdot)$. Depending on the structure of $f(\cdot)$, $h(\cdot)$ may take the form of $h(d)$ for simple parametric models and $h\left(n,d\right)$ for complex nonparametric models, where the best-case scenario is $h\left(d\right)=\mathcal{O}(d)$ for $f(\cdot)$ representing a linear regression model. Hence the time complexity for $2n$ evaluations of $f(\cdot)$ is at least $\mathcal{O}(nd)$ and usually much larger. The main additional computational expense of the PALE total effect VIMs comes from the path finding algorithms. Finding the paths takes $\mathcal{O}(n\log(\frac{n}{K}))$ time for ${ALE}_{j,T}^Q$, and $\mathcal{O}(nd\log(\frac{n}{K}))$ for ${ALE}_{j,T}^C$. \cite{Apley2020} recommends choosing $K$ proportional to $n$, in which case $\log(\frac{n}{K})$ is essentially a constant, and the path finding algorithm takes $\mathcal{O}(nd)$ time, which is smaller than $\mathcal{O}\left(nh(d)\right)$ or $\mathcal{O}(nh\left(n,d\right))$. Hence, the latter can be viewed as the overall time complexity of ${ALE}_{j,T}^Q$ and ${ALE}_{j,T}^C$.

\par
In comparison, the marginal permutation importance ${MP}_j$ based on training error requires $2n$ evaluations of $f(\cdot)$ to calculate the decrease in prediction accuracy for a single permutation of $X_j$. To achieve a more accurate estimate of the mean decrease in prediction accuracy and to reduce sampling variation due to the random nature of permutations, multiple replicates are recommended (see e.g. \cite{fisher2019all} and \cite{molnar2020interpretable}), where each replicate corresponds to a different random permutation of $X_j$. If we use the average increase in loss across $N$ replicates to compute ${MP}_j$, the time complexity is $\mathcal{O}(Nnh\left(d\right))$ or $\mathcal{O}(Nnh\left(n,d\right))$. If one selects $N\gg 2$ ($N$ defaults to $5$ in \textbf{\texttt{iml}} and Python \textbf{scikit-learn} and $N\geq30$ is recommended), this is more expensive than the PALE VIMs. Moreover, if one uses increase in cross-validation or test error loss, instead of training error loss, the computational expense of the permutation VIM is far higher. Shapley VIMs are generally far more computationally expensive than the other VIMs. The \textbf{\texttt{iml}} package that we used in Section \ref{Section6Examples} implements the approximation technique of \cite{vstrumbelj2014}. It requires $2M$ evaluations of $f(\cdot)$ to compute a local Shapley value ($M$ defaults to 100 and larger $M$ values are recommended).  To compute the global ${SHM}_j$, one needs to aggregate local Shapley values across the training sample, resulting in the overall time complexity of $\mathcal{O}(Mnh\left(d\right))$ or $\mathcal{O}(Mnh\left(n,d\right))$.

\par
To make the discussion more concrete, for the examples in Sections \ref{Section6.2NumericalExamples} and \ref{Section6.3BikeSharing} implemented on a Windows™ laptop with Intel(R) Core(TM) i7-1165G7 CPU @ 2.80 GHz processor, ${ALE}_{j,M}$ and ${ALE}_{j,T}^Q$ each take less than one second to compute for each $j$. ${ALE}_{j,T}^C$ takes about 2-3 seconds for Example \ref{example: numerical example 4 predictors interaction} ($n=10,000$, $d=4$) and 3 seconds for Example \ref{example: bikesharing example} ($n=17,333$, $d=10$) for each $j$. The range of times for each example is mainly due to the different time complexities of evaluating $f(\cdot)$ for the different models.  Depending on $n$, $d$, and $f(\cdot)$ in Examples \ref{example: numerical example 4 predictors interaction} and \ref{example: bikesharing example}, the permutation VIMs for each predictor take about 10 seconds to 1 minutes to compute using $N=50$ replicates. The Shapley values take about 12 minutes to 40 minutes to compute for each predictor using the default $M=100$. 

\section{Conclusions}\label{Section7Conclusions}
\par
To assess the importance of predictor variables in black box supervised learning models, sampling from distributions of predictors or subsets of predictors is usually required. While the debate on using marginal versus conditional distributions in such sampling is still ongoing, we have argued that marginal approaches such as permutation VIMs and marginal Shapley values yield more theoretically reasonable results that better represent the importances of the predictors, whereas the conditional versions can be highly misleading (e.g., assigning substantial importance to predictors that are completely absent from the model and undercounting the importance of important predictors, as in Example \ref{example: theoretical VIMs of linear model}). However, a major disadvantage of the marginal versions is that they require the fitted model to extrapolate outside the training data envelope, which can result in unreliable VIMs, whereas the conditional versions avoid or mitigate the extrapolation problem. The ALE and PALE VIMs proposed in this paper enjoy the advantages of both the marginal and conditional VIMs without the disadvantages. When the model $f(\cdot)$ can reliably extrapolate (as in Example \ref{example: theoretical VIMs of linear model}, in which $f(\cdot)$ is known) so that the marginal Shapley and permutation VIMs are reliable, our PALE VIMs produce similar results but with substantially lower computational expense. And when $f(\cdot)$ is unable to reliably extrapolate, the ALE and PALE VIMs are not nearly as adversely affected since they avoid extrapolation and produce more consistent results across replicates. Moreover, the ALE and PALE VIMs possess all of the desirable properties discussed in Section \ref{Section2DesirableProperties}, including much lower computational expense than existing VIMs (especially Shapley VIMs) and producing main effect (and, if desired, second-order interaction effect) VIMs that can be compared to the total effect VIMs to gauge the strengths of interactions and identify which variables interact most with other variables. 
\bibliography{citations_vim}
\bibliographystyle{rss.bst}

\clearpage
\setcounter{page}{1}
\section*{\hfil Supplementary Materials: Appendices\hfil}

\begin{appendices}
\numberwithin{equation}{section}
\renewcommand{\theequation}{\Alph{section}.\arabic{equation}}

\section{Statements and Proofs of Theorems 1 and 2} \label{Appendix: theorems and proofs}
\begin{thm1}
${ALE}_{j,T}^Q\geq{ALE}_{j,M}$ with equality if and only if $f(\cdot)$ is additive in $X_j$, i.e.\ $f\left(\mathbf{x}\right)=f_j\left(x_j\right)+f_{\backslash j}\left(\mathbf{x}_{\backslash j}\right)$, in which case ${ALE}_{j,T}^Q={ALE}_{j,M}=\mathrm{Var}\left[f_j\left(X_j\right)\right]$.  
\end{thm1}
\begin{proof}
Using the standard variance decomposition, we can write 
\begin{equation}\label{eq: proof thm1 var decomposition}
{ALE}_{j,T}^Q=\mathrm{Var}\left[g_j\left(X_j,U;c\right)\right]=\mathbb{E}\left[\mathrm{Var}\left[g_j\left(X_j,U;c\right)|X_j\right]\right]+\mathrm{Var}\left[\mathbb{E}\left[g_j\left(X_j,U;c\right)|X_j\right]\right].
\end{equation}
We also have
\begin{equation}\label{eq: proof thm1 Expectation of g}
\begin{aligned}
&\mathbb{E}\left[g_j\left(X_j,U;c\right)|X_j=x\right]=\int_{0}^{1}{g_j\left(x,u;c\right)du}=\int_{0}^{1}{\int_{c}^{x}{f_u^j\left(z\right)dz}du}	\\
&=\int_{c}^{x}{\int_{0}^{1}{f_u^j\left(z\right)du}dz}=\int_{c}^{x}\mathbb{E}\left[f^j\left(X_j,\mathbf{X}_{\backslash j}\right)|X_j=z\right]dz=f_{j,ALE}\left(x\right)+\mathrm{constant}.
\end{aligned}
\end{equation}
The third equality in \eqref{eq: proof thm1 Expectation of g} follows from Fubini’s Theorem, and the fourth equality follows from the well-known property that the expected value of a random variable is the integral of its quantile function over the interval $[0,1]$. From \eqref{eq: proof thm1 Expectation of g}, we have 
\begin{equation}\label{eq: proof thm1 var equals main VIM}
\mathrm{Var}\left[\mathbb{E}\left[g_j\left(X_j,U;c\right)|X_j\right]\right]=\mathrm{Var}\left[f_{j,ALE}\left(X_j\right)\right]={ALE}_{j,M}.
\end{equation}
Inserting \eqref{eq: proof thm1 var equals main VIM} into \eqref{eq: proof thm1 var decomposition} gives
\begin{equation}\label{eq: proof thm1 VIMQ ge main}
{ALE}_{j,T}^Q=\mathbb{E}\left[\mathrm{Var}\left[g_j\left(X_j,U;c\right)|X_j\right]\right]+{ALE}_{j,M}\geq{ALE}_{j,M}.
\end{equation}
Eq.\ \eqref{eq: proof thm1 VIMQ ge main} is an equality if and only if $\mathbb{E}\left[\mathrm{Var}\left[g_j\left(X_j,U;c\right)|X_j\right]\right]=0$, which, since $\mathrm{Var}\left[g_j\left(X_j,U;c\right)|X_j\right]$ is a non-negative random variable, happens if and only if $\mathrm{Var}\left[g_j\left(X_j,U;c\right)|X_j\right]=0$ a.s.$-p_j$. For any $x\in\mathcal{S}_j$, $\mathrm{Var}\left[g_j\left(X_j,U;c\right)|X_j=x\right]=0$ if and only if $g_j\left(x,u;c\right)$ does not depend on $u$. But $g_j\left(x,u;c\right)=\int_{c}^{x}{f_u^j\left(z\right)dz}$ does not depend on $u$ for any $x\in\mathcal{S}_j$ if and only if $f_u^j\left(x\right)$ does not depend on  $u$, i.e.\ $f^j\left(x_j,\mathbf{x}_{\backslash j}\right)$ does not depend on $\mathbf{x}_{\backslash j}$. The latter condition is the definition of $x_j$ having no interaction with any variables in $\mathbf{x}_{\backslash j}$, i.e.\ $f\left(\mathbf{x}\right)=f_j\left(x_j\right)+f_{\backslash j}\left(\mathbf{x}_{\backslash j}\right)$, in which case $\mathrm{Var}\left[f_{j,ALE}\left(X_j\right)\right]=\mathrm{Var}\left[f_j\left(X_j\right)\right]$ (\cite{Apley2020}).
\end{proof}

\begin{thm2}
${ALE}_{j,T}^S\geq{ALE}_{j,M}$ with equality if and only if $f(\cdot)$ is additive in $X_j$, i.e.\ $f\left(\mathbf{x}\right)=f_j\left(x_j\right)+f_{\backslash j}\left(\mathbf{x}_{\backslash j}\right)$, in which case ${ALE}_{j,T}^S={ALE}_{j,M}=\mathrm{Var}\left[f_j\left(X_j\right)\right]$. 
\end{thm2}

\begin{proof}
Using the standard variance decomposition, we can write 
\begin{equation}\label{eq: proof thm2 var decomposition}
{ALE}_{j,T}^S=\mathrm{Var}\left[g_{j,\mathbf{L}}\left(X_j,\omega;c\right)\right]=\mathbb{E}\left[\mathrm{Var}\left[g_{j,\mathbf{L}}\left(X_j,\omega;c\right)|X_j\right]\right]+\mathrm{Var}\left[\mathbb{E}\left[g_{j,\mathbf{L}}\left(X_j,\omega;c\right)|X_j\right]\right].
\end{equation}
We also have 
\begin{equation}\label{eq: proof thm2 Expectation of g_j,L}
\begin{aligned}
&\mathbb{E}\left[g_{j,\mathbf{L}}\left(X_j,\omega;c\right)|X_j=x\right]=\mathbb{E}\left[\int_{c}^{x}{f^j\left(z,\mathbf{L}_z\left(\omega\right)\right)dz}\right]=\int_{\Omega}{\int_{c}^{x}{f^j\left(z,\mathbf{L}_z\left(\omega\right)\right)dz}dP\left(\omega\right)} \\
&=\int_{c}^{x}{\int_{\Omega}{f^j\left(z,\mathbf{L}_z\left(\omega\right)\right)dP\left(\omega\right)}dz}=\int_{c}^{x}\int{p_{\backslash j|j}\left(\mathbf{x}_{\backslash j}|z\right)f^j\left(z,\mathbf{x}_{\backslash j}\right)d\mathbf{x}_{\backslash j}dz}\\
&=\int_{c}^{x}\mathbb{E}\left[f^j\left(X_j,\mathbf{X}_{\backslash j}\right)|X_j=z\right]dz=f_{j,ALE}\left(x\right)+\mathrm{constant},
\end{aligned}
\end{equation}
where the third equality follows from Fubini's theorem and Property \textit{P2}, and the fourth equality follows from Property \textit{P1}. Using \eqref{eq: proof thm2 Expectation of g_j,L} in \eqref{eq: proof thm2 var decomposition} gives
\begin{equation}\label{eq: proof thm2 VIM_S ge main}
{ALE}_{j,T}^S=\mathbb{E}\left[\mathrm{Var}\left[g_{j,\mathbf{L}}\left(X_j,\omega;c\right)|X_j\right]\right]+\mathrm{Var}\left[f_{j,ALE}\left(X_j\right)\right]\geq \mathrm{Var}\left[f_{j,ALE}\left(X_j\right)\right]={ALE}_{j,M}.
\end{equation}
Eq.\ \eqref{eq: proof thm2 VIM_S ge main} is an equality if and only if $\mathbb{E}\left[\mathrm{Var}\left[g_{j,\mathbf{L}}\left(X_j,\omega;c\right)|X_j\right]\right]=0$, which, since $\mathrm{Var}\left[g_{j,\mathbf{L}}\left(X_j,\omega;c\right)|X_j\right]$ is a nonnegative random variable, happens if and only if $\mathrm{Var}\left[g_{j,\mathbf{L}}\left(X_j,\omega;c\right)|X_j\right]=0$ a.s.$-p_j$. Now for any $x\in\mathcal{S}_j$, $\mathrm{Var}\left[g_{j,\mathbf{L}}\left(X_j,\omega;c\right)|X_j=x\right]=0$ if and only if $g_{j,\mathbf{L}}\left(x,\omega;c\right)=\int_{c}^{x}{f^j\left(z,\mathbf{L}_z\left(\omega\right)\right)dz}$ is constant with respect to $\omega$ a.s.$-P$, which happens for almost all (per Lebesgue measure ) $x\in\mathcal{S}_j$ if and only if for almost all $z\in\mathcal{S}_j$, $f^j\left(z,\mathbf{L}_z\left(\omega\right)\right)$ is constant with respect to $\omega$ a.s.$-P$. The latter condition is equivalent to $f\left(\mathbf{x}\right)=f_j\left(x_j\right)+f_{\backslash j}\left(\mathbf{x}_{\backslash j}\right)$ a.s., in which case $\mathrm{Var}\left[f_{j,ALE}\left(X_j\right)\right]=\mathrm{Var}\left[f_j\left(X_j\right)\right]$.
\end{proof}

\begin{remark} \label{Remark: constant path continued}
Eq.\ \eqref{eq: proof thm2 VIM_S ge main} shows that ${ALE}_{j,T}^S = {ALE}_{j,M} + \mathbb{E}\left[\mathrm{Var}\left[g_{j,\mathbf{L}}\left(X_j,\omega;c\right)|X_j\right]\right]$. Thus, the difference between ${ALE}_{j,T}^S$ and ${ALE}_{j,M}$ increases as $\mathbb{E}\left[\mathrm{Var}\left[g_{j,\mathbf{L}}\left(X_j,\omega;c\right)|X_j\right]\right]$ increases. This seems desirable, since we can view $\mathbb{E}\left[\mathrm{Var}\left[g_{j,\mathbf{L}}\left(X_j,\omega;c\right)|X_j\right]\right]$ as a measure of the strength of interactions between $X_j$ and the other predictors, at least for appropriate choice of paths $\mathbf{L}$. For example, for the special case that $\mathbf{X}_{\backslash j}$ is independent of $X_j$, suppose we define the paths such that $\mathbf{L}_x\left(\omega\right)=\mathbf{T}\left(\omega\right)$ where the mapping $\mathbf{T}:\Omega\longmapsto\mathbb{R}^{d-1}$ does not depend on $x$ and is such that the random vector $\mathbf{T}\left(\omega\right)\sim p_{\backslash j|j}\left(\cdot|x\right)=p_{\backslash j}\left(\cdot\right)$ follows the same distribution as $\mathbf{X}_{\backslash j}$. Since $\mathbf{L}_x\left(\omega\right)$ remains constant as $x\in\mathcal{S}_j$ varies along each path, it follows that $g_{j,\mathbf{L}}\left(x,\omega;c\right)=\int_{c}^{x}{f^j\left(z,\mathbf{T}\left(\omega\right)\right)dz}=f\left(x,\mathbf{T}\left(\omega\right)\right)-f\left(c,\mathbf{T}\left(\omega\right)\right)$, and $\mathbb{E}\left[\mathrm{Var}\left[g_{j,\mathbf{L}}\left(X_j,\omega;c\right)|X_j\right]\right]=\mathbb{E}\left[\mathrm{Var}\left[f\left(X_j,\mathbf{X}_{\backslash j}\right)-f\left(c,\mathbf{X}_{\backslash j}\right)|X_j\right]\right]$. The quantity $f\left(X_j,\mathbf{X}_{\backslash j}\right)-f\left(c,\mathbf{X}_{\backslash j}\right)$ is the change in $f\left(\mathbf{x}\right)$ as $x_j$ is varied from $c$ to $X_j$ with the other predictors held fixed at $\mathbf{x}_{\backslash j}=\mathbf{X}_{\backslash j}$. As the strength of the $X_j$ interactions increases, these changes in $f\left(\mathbf{x}\right)$ will depend more strongly on $\mathbf{X}_{\backslash j}$, in which case we would expect $\mathbb{E}\left[\mathrm{Var}\left[f\left(X_j,\mathbf{X}_{\backslash j}\right)-f\left(c,\mathbf{X}_{\backslash j}\right)|X_j\right]\right]={ALE}_{j,T}^S-{ALE}_{j,M}$ to increase. More generally, when $\mathbf{X}_{\backslash j}$ is not independent of $X_j$, this choice of constant paths violates property \textit{P1} and would also require extrapolating $f\left(\mathbf{x}\right)$. Our specific choice of connected paths discussed in Section \ref{Section5.4ConnectedSPALE} attempts to keep the paths as constant as possible while satisfying properties \textit{P1} and \textit{P2} and avoiding extrapolation. 
\end{remark}

\section{Additional Details of Computing ${\widehat{ALE}}_{j,T}^Q$ and ${\widehat{ALE}}_{j,T}^S$}\label{Appendix: sample versions of PALE VIMs}
\par
\textbf{Computing ${\widehat{ALE}}_{j,T}^Q$.} After computing the functions ${\hat{g}}_j\left(x,u;z_{0,j}\right)$ for each $x\in\left\{z_{k,j}: k=1, 2,\ldots, K\right\}$ and $u\in\left\{u_l=\left(l-1/2\right)/L: l=1,2, \ldots, L\right\}$ as described in Section \ref{Section5.2QPALE} (Eq.\ \eqref{eq: estimator of QPALE function from z0j to x}), we find the sample version of $c^\ast$ as follows.  
To simplify the calculation, use the fact that 
\begin{equation}\label{eq: QPALE from c to x as difference of two g}
g_j\left(x,u;c\right)=g_j\left(x,u;z_{0,j}\right)-g_j\left(c,u;z_{0,j}\right),
\end{equation}
which gives 
\begin{equation}\label{eq: var of QPALE to be minimized}
\begin{aligned}
&\mathrm{Var}\left[g_j\left(X_j,U;c\right)\right]=\mathrm{Var}\left[\mathbb{E}\left[g_j\left(X_j,U;c\right)|U\right]\right]+\mathbb{E}\left[\mathrm{Var}\left[g_j\left(X_j,U;c\right)|U\right]\right] \\
& =\mathrm{Var}\left[\mathbb{E}\left[g_j\left(X_j,U;z_{0,j}\right)-g_j\left(c,U;z_{0,j}\right)|U\right]\right]+\mathbb{E}\left[\mathrm{Var}\left[g_j\left(X_j,U;z_{0,j}\right)-g_j\left(c,U;z_{0,j}\right)|U\right]\right] \\
& =\mathrm{Var}\left[{\bar{g}}_j\left(\cdot,U;z_{0,j}\right)-g_j\left(c,U;z_{0,j}\right)\right]+\mathbb{E}\left[\mathrm{Var}\left[g_j\left(X_j,U;z_{0,j}\right)|U\right]\right],
\end{aligned}
\end{equation}
where
\begin{equation}\label{eq: QPALE g_j_bar dot u}
{\bar{g}}_j\left(\cdot,U;z_{0,j}\right)=\mathbb{E}\left[g_j\left(X_j,U;z_{0,j}\right)|U\right].
\end{equation}
Because the right-most term in \eqref{eq: var of QPALE to be minimized} does not depend on $c$, the value $c^\ast$ that minimizes $\mathrm{Var}\left[g_j\left(X_j,U;c\right)\right]$ also minimizes
\begin{equation}\label{eq: var QPALE to be minimized numerically}
\mathrm{Var}\left[{\bar{g}}_j\left(\cdot,U;z_{0,j}\right)-g_j\left(c,U;z_{0,j}\right)\right]=\int_{0}^{1}{\left[\left({\bar{g}}_j\left(\cdot,u;z_{0,j}\right)-{\bar{g}}_j\left(\cdot,\cdot;z_{0,j}\right)\right)-\left(g_j\left(c,u;z_{0,j}\right)-{\bar{g}}_j\left(c,\cdot;z_{0,j}\right)\right)\right]^2 du},
\end{equation}
where
\begin{equation}\label{eq: QPALE g bar dot dot grand mean}
{\bar{g}}_j\left(\cdot,\cdot;z_{0,j}\right)=\mathbb{E}\left[g_j\left(X_j,U;z_{0,j}\right)\right]=\mathbb{E}\left[{\bar{g}}_j\left(\cdot,U;z_{0,j}\right)\right], 
\end{equation} and
\begin{equation}\label{eq: QPALE g bar c dot}
{\bar{g}}_j\left(c,\cdot;z_{0,j}\right)=\mathbb{E}\left[g_j\left(X_j,U;z_{0,j}\right)|X_j=c\right]. 
\end{equation}
Thus, to minimize the sample version of \eqref{eq: var QPALE to be minimized numerically}, we take the sample version of $c^\ast$ to be $z_{k^\ast,j}$, where 
\begin{equation}\label{eq: QPALE k star estimator}
k^\ast=\argmin_k\sum_{l=1}^{L}\left[\left({\bar{\hat{g}}}_j\left(\cdot,u_l;z_{0,j}\right)-{\bar{\hat{g}}}_j\left(\cdot,\cdot;z_{0,j}\right)\right)-\left({\hat{g}}_j\left(z_{k,j},u_l;z_{0,j}\right)-{\bar{\hat{g}}}_j\left(z_{k,j},\cdot;z_{0,j}\right)\right)\right]^2,
\end{equation}
and, in analogy with \eqref{eq: QPALE g_j_bar dot u}, \eqref{eq: QPALE g bar c dot}, and \eqref{eq: QPALE g bar dot dot grand mean}, 
\begin{equation}\label{eq: estimator QPALE g_j_bar dot u}
{\bar{\hat{g}}}_j\left(\cdot,u_l;z_{0,j}\right)=\sum_{k=1}^{K}{\frac{n_j\left(k\right)}{n}{\hat{g}}_j\left(z_{k,j},u_l;z_{0,j}\right)}, 
\end{equation}

\begin{equation}\label{eq: estimator QPALE g bar c dot}
{\bar{\hat{g}}}_j\left(z_{k,j},\cdot;z_{0,j}\right)=\sum_{l=1}^{L}{\frac{1}{L}{\hat{g}}_j\left(z_{k,j},u_l;z_{0,j}\right)}, 
\end{equation} and
\begin{equation}\label{eq: estimator QPALE g bar dot dot grand mean}
{\bar{\hat{g}}}_j\left(\cdot,\cdot;z_{0,j}\right)=\sum_{k=1}^{K}\sum_{l=1}^{L}{\frac{n_j\left(k\right)}{nL}{\hat{g}}_j\left(z_{k,j},u_l;z_{0,j}\right)}=\sum_{k=1}^{K}{\frac{n_j\left(k\right)}{n}{\bar{\hat{g}}}_j\left(z_{k,j},\cdot;z_{0,j}\right)}=\sum_{l=1}^{L}{\frac{1}{L}{\bar{\hat{g}}}_j\left(\cdot,u_l;z_{0,j}\right)}.
\end{equation}
Finally, ${\widehat{ALE}}_{j,T}^Q$ is the sample variance of ${\hat{g}}_j\left(x_{i,j},u_l;z_{k^\ast,j}\right)={\hat{g}}_j\left(x_{i,j},u_l;z_{0,j}\right)-{\hat{g}}_j\left(z_{k^\ast,j},u_l;z_{0,j}\right)$ across $\{(x_{i,j}, u_l): i = 1, 2, \ldots, n; l = 1, 2, \ldots, L\}$, i.e., 

\begin{equation}\label{eq: estimator QPALE VIM}
{\widehat{ALE}}_{j,T}^Q=\sum_{k=1}^{K}\sum_{l=1}^{L}{\frac{n_j\left(k\right)}{nL}\left[\left({\hat{g}}_j\left(z_{k,j},u_l;z_{0,j}\right)-{\hat{g}}_j\left(z_{k^\ast,j},u_l;z_{0,j}\right)\right)-\left({\bar{\hat{g}}}_j\left(\cdot,\cdot;z_{0,j}\right)-{\bar{\hat{g}}}_j\left(z_{k^\ast,j},\cdot;z_{0,j}\right)\right)\right]^2}.
\end{equation}

\par
For notational simplicity, in  \eqref{eq: estimator of QPALE function from z0j to x} and the preceding derivations, we have used the piecewise constant representation ${\hat{g}}_j\left(x_{i,j},u_l;z_{0,j}\right)={\hat{g}}_j\left(z_{k_j\left(x_{i,j}\right),j},u_l;z_{0,j}\right)$, i.e., ${\hat{g}}_j\left(x,u_l;z_{0,j}\right)$ assumes the same constant value across the interval $x\in N_j\left(k\right)$, which makes computation of $k^\ast$ more efficient. In our coded implementation (included in the supplementary materials), we also use a piecewise constant representation but take the constant value across the interval $x \in N_j(k)$ to be $\frac{1}{2}\left({\hat{g}}_j\left(z_{k-1, j},u_l;z_{0,j}\right) + {\hat{g}}_j\left(z_{k,j},u_l;z_{0,j}\right)\right)$, i.e, the average of the values at the left and right endpoints of the interval. Alternatively, one could  interpolate ${\hat{g}}_j\left(x_{i,j},u_l;z_{0,j}\right)$ linearly for $x_{i,j}\in\left[z_{k_j\left(x_{i,j}\right)-1,j},z_{k_j\left(x_{i,j}\right),j}\right]$, which would only minimally increase computational expense. To compute ${\widehat{ALE}}_{j,T}^Q$, we must compute the $n$ individual local effects $\{f\left(z_{k_j\left(x_{i,j}\right),j},\ \mathbf{x}_{i,\backslash j}\right)-f\left(z_{k_j\left(x_{i,j}\right)-1,j}, \mathbf{x}_{i,\backslash j}\right):i=1,2,\ldots,n\}$, which requires only $2n$ evaluations of $f(\cdot)$, regardless of $K$. From these, we compute the $K\times L\approx n$ function values $\left\{{\hat{g}}_j\left(z_{k,j},u_l;z_{0,j}\right):k=1,2,\ldots,K;l=1,2,\ldots,L\right\}$, from which all of the functions involved in \eqref{eq: QPALE k star estimator} to \eqref{eq: estimator QPALE VIM} can be efficiently computed. 

\par
\textbf{Computing ${\widehat{ALE}}_{j,T}^S$.} After computing the functions ${\hat{g}}_{j,L}\left(x,\omega_l;z_{0,j}\right)$ for each $x\in\left\{z_{k,j}: k=1, 2, \ldots, K\right\}$ and for each $l=1, 2,\ldots, L$ as described in Section \ref{Section5.3SPALE} (Eq.\ \eqref{eq: estimator of SPALE function from z0j to x}), we must find the sample version of $c^\ast$ for this formulation. A straightforward repetition of the derivations in Eqs.\ \eqref{eq: QPALE from c to x as difference of two g} to \eqref{eq: QPALE g bar c dot} (with $U\sim\mathrm{Uniform}\left[0,1\right]$ replaced by $\omega\sim P$) reveals that the value $c$ that minimizes $\mathrm{Var}\left[g_{j,\mathbf{L}}\left(X_j,\omega;c\right)\right]$ also minimizes
\begin{equation}\label{eq: var SPALE to be minimized numerically}
\begin{aligned}
&\mathrm{Var}\left[{\bar{g}}_{j,\mathbf{L}}\left(\cdot,\omega;z_{0,j}\right)-g_{j,\mathbf{L}}\left(c,\omega;z_{0,j}\right)\right]\\
&=\int_{\Omega}{\left[\left({\bar{g}}_{j,\mathbf{L}}\left(\cdot,\omega;z_{0,j}\right)-{\bar{g}}_{j,\mathbf{L}}\left(\cdot,\cdot;z_{0,j}\right)\right)-\left(g_{j,\mathbf{L}}\left(c,\omega;z_{0,j}\right)-{\bar{g}}_{j,\mathbf{L}}\left(c,\cdot;z_{0,j}\right)\right)\right]^2 dP\left(\omega\right)},
\end{aligned}
\end{equation}
where
\begin{equation}\label{eq: SPALE g_j_bar dot omega}
{\bar{g}}_{j,\mathbf{L}}\left(\cdot,\omega;z_{0,j}\right)=\mathbb{E}\left[g_{j,\mathbf{L}}\left(X_j,\omega;z_{0,j}\right)|\omega\right],   
\end{equation}

\begin{equation}\label{eq: SPALE g bar dot dot grand mean}
{\bar{g}}_{j,\mathbf{L}}\left(\cdot,\cdot;z_{0,j}\right)=\mathbb{E}\left[g_{j,\mathbf{L}}\left(X_j,\omega;z_{0,j}\right)\right]=\mathbb{E}\left[{\bar{g}}_{j,\mathbf{L}}\left(\cdot,\omega;z_{0,j}\right)\right], 
\end{equation}
and
\begin{equation}\label{eq: SPALE g bar c dot}
{\bar{g}}_{j,\mathbf{L}}\left(c,\cdot;z_{0,j}\right)=\mathbb{E}\left[g_{j,\mathbf{L}}\left(X_j,\omega;z_{0,j}\right)|X_j=c\right]. 
\end{equation}
To minimize the sample version of \eqref{eq: var SPALE to be minimized numerically}, we take the sample version of $c^\ast$ to be $z_{k^\ast,j}$, where 
\begin{equation}\label{eq: SPALE k star estimator}
k^\ast=\argmin_k\sum_{l=1}^{L}\left[\left({\bar{\hat{g}}}_{j,\mathbf{L}}\left(\cdot,\omega_l;z_{0,j}\right)-{\bar{\hat{g}}}_{j,\mathbf{L}}\left(\cdot,\cdot;z_{0,j}\right)\right)-\left({\hat{g}}_{j,\mathbf{L}}\left(z_{k,j},\omega_l;z_{0,j}\right)-{\bar{\hat{g}}}_{j,\mathbf{L}}\left(z_{k,j},\cdot;z_{0,j}\right)\right)\right]^2,
\end{equation}
\begin{equation}\label{eq: estimator SPALE g_j_bar dot omega}
{\bar{\hat{g}}}_{j,\mathbf{L}}\left(\cdot,\omega_l;z_{0,j}\right)=\sum_{k=1}^{K}{\frac{n_j\left(k\right)}{n}{\hat{g}}_{j,\mathbf{L}}\left(z_{k,j},\omega_l;z_{0,j}\right)}, 
\end{equation}

\begin{equation}\label{eq: estimator SPALE g bar c dot}
{\bar{\hat{g}}}_{j,\mathbf{L}}\left(z_{k,j},\cdot;z_{0,j}\right)=\sum_{l=1}^{L}{\frac{1}{L}{\hat{g}}_{j,\mathbf{L}}\left(z_{k,j},\omega_l;z_{0,j}\right)}, 
\end{equation}
and
\begin{equation}\label{eq: estimator SPALE g bar dot dot grand mean}
{\bar{\hat{g}}}_{j,\mathbf{L}}\left(\cdot,\cdot;z_{0,j}\right)=\sum_{k=1}^{K}\sum_{l=1}^{L}{\frac{n_j\left(k\right)}{nL}{\hat{g}}_{j,\mathbf{L}}\left(z_{k,j},\omega_l;z_{0,j}\right)}=\sum_{k=1}^{K}{\frac{n_j\left(k\right)}{n}{\bar{\hat{g}}}_{j,\mathbf{L}}\left(z_{k,j},\cdot;z_{0,j}\right)}=\sum_{l=1}^{L}{\frac{1}{L}{\bar{\hat{g}}}_{j,\mathbf{L}}\left(\cdot,\omega_l;z_{0,j}\right)}.   
\end{equation}

Finally, ${\widehat{ALE}}_{j,T}^S$ is the sample variance of ${\hat{g}}_{j,\mathbf{L}}\left(x_{i,j},\omega_l;z_{k^\ast,j}\right)={\hat{g}}_{j,\mathbf{L}}\left(x_{i,j},\omega_l;z_{0,j}\right)-{\hat{g}}_{j,\mathbf{L}}\left(z_{k^\ast,j},\omega_l;z_{0,j}\right)$ across $\left\{(x_{i,j}, \omega_l ): i = 1, 2, \ldots, n; l = 1, 2, \ldots, L\right\}$, i.e., 

\begin{equation}\label{eq: estimator SPALE VIM}
{\widehat{ALE}}_{j,T}^S=\sum_{k=1}^{K}\sum_{l=1}^{L}{\frac{n_j\left(k\right)}{nL}\left[\left({\hat{g}}_{j,\mathbf{L}}\left(z_{k,j},\omega_l;z_{0,j}\right)-{\hat{g}}_{j,\mathbf{L}}\left(z_{k^\ast,j},\omega_l;z_{0,j}\right)\right)-\left({\bar{\hat{g}}}_{j,\mathbf{L}}\left(\cdot,\cdot;z_{0,j}\right)-{\bar{\hat{g}}}_{j,\mathbf{L}}\left(z_{k^\ast,j},\cdot;z_{0,j}\right)\right)\right]^2}.
\end{equation}

\section{PALE VIM Definitions for Nondifferentiable $f(\cdot)$}\label{Appendix: PALE definitions nondifferentiable f}
\par 
\cite{Apley2020} defined ALE functions for nondifferentiable $f(\cdot)$ and showed that this generalization reduces to the definitions \eqref{eq: ALE main uncentered} and \eqref{eq: ALE main centered} if $f(\cdot)$ is differentiable. In this appendix, we use the same concepts to briefly describe how to generalize the definition of PALE total effect VIMs to non-differentiable $f(\cdot)$. It is important to note that the sample versions of PALE VIMs are the same regardless of whether $f(\cdot)$ is differentiable, since the sample versions use finite differences and not derivatives. 

\par
To simplify the presentation, consider a continuous numerical predictor $X_j$ with support $\mathcal{S}_j=\left[x_{\min,j},x_{\max,j}\right]$. For $K=1,2,\ldots$, let $\mathcal{P}_j^K\equiv\{z_{k,j}^K:k=0, 1,\ldots,K\}$ represent a partition of $\mathcal{S}_j$ into $K$ intervals $\{(z_{k-1,j}^K,z_{k,j}^K]: k=1, 2, \ldots, K\}$ with $z_{0,j}^K=x_{\min,j}$ and $z_{K,j}^K=x_{\max,j}$, and let $\delta_{j,K}\equiv\max\{|z_{k,j}^K-z_{k-1,j}^K|:k=1, 2,\ldots,K\}$ denote the fineness of the partition. For any $x\in\mathcal{S}_j$, we use $k_j^K(x)$ to denote the index of the interval of $\mathcal{P}_j^K$ into which $x$ falls, i.e., $x\in (z_{k-1,j}^K,z_{k,j}^K]$ for $k=k_j^K(x)$. Suppose the sequence of partitions $\{\mathcal{P}_j^K:K=1, 2, \ldots\}$ is such that $\lim_{K\rightarrow\infty}{\delta_{j,K}}=0$. 

\par
For each $k=1,2,\ldots,K$ and $u\in\left[0,1\right]$, let $\Delta_j^K\left(k,u\right)$ denote the $u$-quantile of the conditional distribution of $f\left(z_{k,j}^K,\mathbf{X}_{\backslash j}\right)-f\left(z_{k-1,j}^K,\mathbf{X}_{\backslash j}\right)|X_j\in(z_{k-1,j}^K,z_{k,j}^K]$, i.e., such that $\mathbb{P}(f\left(z_{k,j}^K, \mathbf{X}_{\backslash j}\right)-f\left(z_{k-1,j}^K, \mathbf{X}_{\backslash j}\right)\le\Delta_j^K\left(k,u\right)|X_j\in(z_{k-1,j}^K,z_{k,j}^K])=u$. For each $\left(x, u\right)\in\mathcal{S}_j\times[0,1]$ and some constant $c\in\mathcal{S}_j$, the generalized QPALE function of $X_j$ is defined as 
\begin{equation}\label{eq: nondifferentiable QPALE function c to x}
g_j\left(x,u;c\right)=\lim_{K\rightarrow\infty}\sum_{k=k_j^K\left(c\right)}^{k_j^K\left(x\right)}{\Delta_j^K\left(k,u\right)}.
\end{equation}
Using \eqref{eq: nondifferentiable QPALE function c to x}, we define the QPALE total effect VIM for $X_j$ analogous to \eqref{eq: QPALE VIM definition} as
\begin{equation}\label{eq: nondifferentiable QPALE VIM definition}
{ALE}_{j,T}^Q=\mathrm{Var}\left[g_j\left(X_j,U;c\right)\right],
\end{equation}
where $U\sim \mathrm{Uniform}[0,1]$ is independent of $X_j$ and the constant $c$ can be chosen as either $\mathbb{E}[X_j]$ or as $c^\ast=\argmin_ c{\mathrm{Var}\left[g_j\left(X_j,U;c\right)\right]}$.

\par
Analogous to Theorem 1, the following theorem shows that ${ALE}_{j,T}^Q$ defined via \eqref{eq: nondifferentiable QPALE function c to x} and \eqref{eq: nondifferentiable QPALE VIM definition} also satisfies desirable properties (\ref{property: total>main}) and (\ref{property: additive recovery}), assuming the functions $\sum_{k=k_j^K\left(c\right)}^{k_j^K\left(x\right)}{\Delta_j^K\left(k,u\right)}$ are bounded in magnitude by some finite constant for all $x$, $u$, $c$, and $K$ (to allow the bounded convergence theorem to be used in the proof below). 
\begin{thmA.1}
${ALE}_{j,T}^Q\geq{ALE}_{j,M}$ with equality if and only if $f(\cdot)$ is additive in $X_j$, i.e.\ $f\left(\mathbf{x}\right)=f_j\left(x_j\right)+f_{\backslash j}\left(\mathbf{x}_{\backslash j}\right)$, in which case ${ALE}_{j,T}^Q={ALE}_{j,M}=\mathrm{Var}\left[f_j\left(X_j\right)\right]$.
\end{thmA.1}
\begin{proof}
Using the standard variance decomposition, we can write 
\begin{equation}\label{eq: nondifferentiable proof thmA1 var decomposition}
{ALE}_{j,T}^Q=\mathrm{Var}\left[g_j\left(X_j,U;c\right)\right]=\mathbb{E}\left[\mathrm{Var}\left[g_j\left(X_j,U;c\right)|X_j\right]\right]+\mathrm{Var}\left[\mathbb{E}\left[g_j\left(X_j,U;c\right)|X_j\right]\right].
\end{equation}
We also have
\begin{equation}\label{eq: nondifferentiable proof thmA1 Expectation of g}
\begin{aligned}
&\mathbb{E}\left[g_j\left(X_j,U;c\right)|X_j=x\right]=\int_{0}^{1}{g_j\left(x,u;c\right)du}=\int_{0}^{1}{\lim_{K\rightarrow\infty}\sum_{k=k_j^K\left(c\right)}^{k_j^K\left(x\right)}{\Delta_j^K\left(k,u\right)}du}\\
& =\lim_{K\rightarrow\infty}\sum_{k=k_j^K\left(c\right)}^{k_j^K\left(x\right)}\int_{0}^{1}{\Delta_j^K\left(k,u\right)du}=\lim_{K\rightarrow\infty}\sum_{k=k_j^K\left(c\right)}^{k_j^K\left(x\right)}\mathbb{E}[f\left(z_{k,j}^K,\mathbf{X}_{\backslash j}\right)-f\left(z_{k-1,j}^K,\mathbf{X}_{\backslash j}\right)|X_j\in(z_{k-1,j}^K,z_{k,j}^K]]\\
&=f_{j,ALE}\left(x\right)+\mathrm{constant}.
\end{aligned}
\end{equation}
The third equality in \eqref{eq: nondifferentiable proof thmA1 Expectation of g} follows from bounded convergence theorem, and the fourth equality follows from the well-known property that the expected value of a random variable is the integral of its quantile function over the interval $[0,1]$. The last equality in \eqref{eq: nondifferentiable proof thmA1 Expectation of g} follows trivially from the generalized definition of ALE main effect in \cite{Apley2020}. From \eqref{eq: nondifferentiable proof thmA1 Expectation of g}, we have 
\begin{equation}\label{eq: nondifferentiable proof thmA1 var equals main VIM}
\mathrm{Var}\left[\mathbb{E}\left[g_j\left(X_j,U;c\right)|X_j\right]\right]=\mathrm{Var}\left[f_{j,ALE}\left(X_j\right)\right]={ALE}_{j,M},
\end{equation}
so that 
\begin{equation}\label{eq: nondifferentiable proof thmA1 VIMQ ge main}
{ALE}_{j,T}^Q=\mathbb{E}\left[\mathrm{Var}\left[g_j\left(X_j,U;c\right)|X_j\right]\right]+{ALE}_{j,M}\geq{ALE}_{j,M}.
\end{equation}
The remainder of the proof is the same as in the proof of Theorem 1.
\end{proof}

\par
One could similarly generalize the definition of the ${ALE}_{j,T}^S$ VIMs to nondifferentiable $f(\cdot)$. However, we do not pursue this here, because it would be mathematically tedious and because the sample versions of ${ALE}_{j,T}^S$ would be exactly the same as the sample versions we defined in the context of differentiable $f(\cdot)$.

\section{Handling Categorical Predictors}\label{Appendix: handling categorical predictors}
\par
To define ALE and PALE VIMs for a categorical predictor $X_j$, we use the same approach that \cite{Apley2020} used to define ALE functions for categorical predictors. The first step is to reorder the levels of the categorical $X_j$ according to how dissimilar the sample values of $\{\mathbf{x}_{i,\backslash j}:i=1, 2,\ldots,n\}$ are across the levels of $X_j$ using a multidimensional-scaling-based method. This reordering helps ensure that extrapolation of $f(\cdot)$ is avoided as much as possible when calculating individual local effects across two neighboring levels. The partition of the sample range of $X_j$ is then taken to coincide with the reordered levels of $X_j$, and $K$ is always set as one less than the number of levels of $X_j$ with the left and right endpoints of each interval taken to be the two neighboring $X_j$ levels. We take $n_j(k)$  to be the total number of observations in the two neighboring levels that form the $k$th interval, and our default value for $L$ is $\min(\frac{1}{K}\sum_{k=1}^{K}n_j(k), 256)$. Based on this partition, the same quantities defined for numeric predictors, e.g. in \eqref{eq: estimator of ALE main uncentered} and \eqref{eq: estimator of ALE main centered}, are used as sample versions of the ALE functions in \eqref{eq: ALE main uncentered} and \eqref{eq: ALE main centered}. Sample versions of the VIMs ${ALE}_{j,M}$ and ${ALE}_{j,2}$ are then computed as in Section \ref{Section4.2ALEMainSecondVIMs}.

\par
To compute the PALE VIMs for $X_j$, we use the same partition. The procedure for computing ${\widehat{ALE}}_{j,T}^Q$ and ${\widehat{ALE}}_{j,T}^C$ for numeric predictors in Sections \ref{Section5.2QPALE} to \ref{Section5.4ConnectedSPALE} can still be used with the following consideration. For categorical predictors, since the intervals coincide with the reordered levels of $X_j$, the number of observations in each interval will vary, perhaps substantially. Consequently, some intervals will have more than $L$ observations, and others will have fewer. This does not affect the computation of ${\widehat{ALE}}_{j,T}^Q$, since the quantile $\Delta_j\left(k,u_l\right)$ is well-defined regardless of the number of observations in the interval. For ${\widehat{ALE}}_{j,T}^C$, we account for this in the path finding Algorithm 1 as follows. For intervals with $n_j\left(k\right)>L$, some of the $L$ leaf regions in the $\mathbf{X}_{\backslash j}$ space for that interval will contain multiple observations, in which case we take the corresponding $\Delta_{j,\mathbf{L}}\left(k,\omega_l\right)$ in Eq.\ \eqref{eq: estimator of SPALE function from z0j to x} to be the average $\Delta_{j,\mathbf{L}}\left(k,\omega_l\right)$ value across all observations in that leaf region. Conversely, for intervals with $n_j\left(k\right)<L$, there will be fewer than $L$ leaf regions in the $\mathbf{X}_{\backslash j}$ space for that interval, since the algorithm does not further split any leaf node that contains a single observation. In this case, the $\Delta_{j,\mathbf{L}}\left(k,\omega_l\right)$ value for any leaf node whose splitting was terminated early because of this premature depletion is used for multiple paths, where the multiple paths are those associated with all corresponding child leaf nodes in the other intervals. 

\par
Finally, when growing the regression tree to partition the $\mathbf{X}_{\backslash j}$ space via Algorithm 1, a categorical predictor $X_m$ in $\mathbf{X}_{\backslash j}$ is handled similarly to how categorical predictors are normally handled when fitting regression trees. When choosing each split, we sort the levels of $X_m$ by the average individual local effect values within each level, and the potential split point is the median of $X_m$ based on this reordering of its levels.

\section{Derivations of the VIMs for Example \ref{example: theoretical VIMs of linear model}.}\label{Appendix: derivations of example theoretical linear}
\par
For Example \ref{example: theoretical VIMs of linear model}, $d=3$, $\left(X_1, X_2\right)$ follows a bivariate standard normal distribution with correlation coefficient $\rho$, $X_3$ follows a standard normal distribution independent of $\left(X_1, X_2\right)$, and $f\left(\mathbf{X}\right)=\beta_1 X_1+\beta_2 X_2+\beta_3 X_3$.

\par
The variance-based GSA total and main effect indices for $X_1$ are \\${GS}_{1,T}=\mathbb{E}\left[\mathrm{Var}\left[\beta_1 X_1+\beta_2 X_2+\beta_3 X_3|X_2,X_3\right]\right]=\beta_1^2\left(1-\rho^2\right)$ and ${GS}_{1,M}=\mathrm{Var}\left[\mathbb{E}\left[\beta_1 X_1+\beta_2 X_2|X_1\right]\right]=\left(\beta_1+\rho\beta_2\right)^2$, respectively. Similarly, the VIMs for $X_2$ and $X_3$ are ${GS}_{2,T}=\beta_2^2\left(1-\rho^2\right)$, ${GS}_{2,M}=\left(\rho\beta_1+\beta_2\right)^2$, and ${GS}_{3,T}={GS}_{3,M}=\beta_3^2$. 

\par
For the conditional Shapley VIMs with $f_u\left(\mathbf{x}_u\right)=\mathbb{E}[f(\mathbf{x}_u,\mathbf{X}_{\backslash u})|\mathbf{X}_u=\mathbf{x}_u]$, we have

\begin{equation*}
\begin{aligned}
f_\emptyset\left(\mathbf{x}_\emptyset\right)&=\mathbb{E}\left[f\left(\mathbf{X}\right)\right]=0,\\ 
f_1\left(x_1\right)&=\left(\beta_1+\rho\beta_2\right)x_1, \\
f_2\left(x_2\right)&=\left(\rho\beta_1+\beta_2\right)x_2, \\
f_3\left(x_3\right)&=\beta_3x_3,\\
f_{1,2}\left(x_1,x_2\right)&=\beta_1 x_1+\beta_2 x_2, \\
f_{1,3}\left(x_1,x_3\right)&=\left(\beta_1+\rho\beta_2\right)x_1+\beta_3 x_3,\\
f_{2,3}\left(x_2,x_3\right)&=\left(\rho\beta_1+\beta_2\right)x_2+\beta_3 x_3, \mathrm{and}\\
f_{1,2,3}\left(x_1,x_2,x_3\right)&=\beta_1 x_1+\beta_2 x_2+\beta_3 x_3,  
\end{aligned}
\end{equation*}
in which case $\phi_1\left(\mathbf{x}\right)=\left(\beta_1+\frac{\rho\beta_2}{2}\right)x_1-\frac{\rho\beta_1}{2}x_2$, $\phi_2\left(\mathbf{x}\right)=\left(\beta_2+\frac{\rho\beta_1}{2}\right)x_2-\frac{\rho\beta_2}{2}x_1$, and $\phi_3\left(\mathbf{x}\right)=\beta_3 x_3$. If we take the global VIM for $X_j$ to be $\mathbb{E}\left[\phi_j^2\left(\mathbf{X}\right)\right]$, then the global conditional Shapley VIMs are ${SHC}_1=\mathbb{E}\left[\phi_1^2\left(\mathbf{X}\right)\right]=\left(\beta_1+\frac{\rho\beta_2}{2}\right)^2+\left(\frac{\rho\beta_1}{2}\right)^2-\rho^2\beta_1\left(\beta_1+\frac{\rho\beta_2}{2}\right)$, ${SHC}_2=\mathbb{E}\left[\phi_2^2\left(\mathbf{X}\right)\right]=\left(\beta_2+\frac{\rho\beta_1}{2}\right)^2+\left(\frac{\rho\beta_2}{2}\right)^2-\rho^2\beta_2\left(\beta_2+\frac{\rho\beta_1}{2}\right)$, and ${SHC}_3=\mathbb{E}\left[\phi_3^2\left(\mathbf{X}\right)\right]=\beta_3^2$.

\par
For the marginal Shapley VIMs with $f_u\left(\mathbf{x}_u\right)=\mathbb{E}[f(\mathbf{x}_u,\mathbf{X}_{\backslash u})]$, we have
\begin{equation*}
\begin{aligned}
f_\emptyset\left(\mathbf{x}_\emptyset\right)&=\mathbb{E}\left[f\left(\mathbf{X}\right)\right]=0,\\
f_1\left(x_1\right)&=\beta_1 x_1,\\
f_2\left(x_2\right)&=\beta_2 x_2,\\
f_3\left(x_3\right)&=\beta_3 x_3,\\
f_{1,2}\left(x_1,x_2\right)&=\beta_1 x_1+\beta_2 x_2,\\ 
f_{1,3}\left(x_1,x_3\right)&=\beta_1 x_1+\beta_3 x_3,\\
f_{2,3}\left(x_2,x_3\right)&=\beta_2 x_2+\beta_3 x_3, \mathrm{and}\\
f_{1,2,3}\left(x_1,x_2,x_3\right)&=\beta_1 x_1+\beta_2 x_2+\beta_3 x_3,
\end{aligned}
\end{equation*}
in which case $\phi_j\left(\mathbf{x}\right)=\beta_j x_j$, and the global marginal Shapley VIMs are ${SHM}_j=\mathbb{E}\left[\phi_j^2\left(\mathbf{X}\right)\right]=\beta_j^2$.

\par
To compute marginal permutation VIMs, consider a response variable $Y=f\left(\mathbf{X}\right) + \varepsilon$ with $\varepsilon\sim N\left(0,\sigma^2\right)$ and independent of $\mathbf{X}$. We use the standard convention and take the VIM for $X_j$ to be the expected increase in squared error loss $\mathbb{E}\left[\left(Y-f\left(\mathbf{X}\right)\right)^2\right]$ when $X_j$ is drawn randomly from marginal distribution $p_j(\cdot)$, which corresponds to marginally permuting $X_j$. Let $X_j^\pi$ denote the permuted version of $X_j$. The marginal permutation VIM for $X_1$ is ${MP}_1=\mathbb{E}\left[\left(Y-\beta_1 X_1^\pi-\beta_2 X_2-\beta_3 X_3\right)^2-\left(Y-\beta_1 X_1-\beta_2 X_2-\beta_3 X_3\right)^2\right]=\mathbb{E}\left[\left(\varepsilon+\beta_1\left(X_1-X_1^\pi\right)\right)^2-\varepsilon^2\right]=2\beta_1^2$. Similarly, the VIMs for $X_2$ and $X_3$ are ${MP}_2=2\beta_2^2$ and ${MP}_3=2\beta_3^2$, respectively.

\par
The conditional permutation VIMs are the same as the marginal permutation VIMs, except we take the VIM for $X_j$ to be the expected increase in squared error loss when $X_j$ is randomly drawn from its conditional distribution $p_{j|\backslash j}(\cdot|\cdot)$, which corresponds to conditionally permuting $X_j$. Let $X_j^C$ denote the conditionally permuted version of $X_j$, which for convenience we can represent as $X_1^C=\rho X_2+\left(1-\rho^2\right)^{1/2}\nu$, $X_2^C=\rho X_1+\left(1-\rho^2\right)^{1/2}\nu$, and $X_3^C=\nu$, where $\nu$ is a standard normal random variable independent of all other variables. The conditional permutation VIM for $X_1$ is
\begin{equation*}
\begin{aligned}
    {CP}_1&=\mathbb{E}\left[\left(Y-\beta_1 X_1^C-\beta_2 X_2-\beta_3 X_3\right)^2-\left(Y-\beta_1 X_1-\beta_2 X_2-\beta_3 X_3\right)^2\right]\\
    &=\mathbb{E}\left[\left(\varepsilon+\beta_1\left(X_1-X_1^C\right)\right)^2-\varepsilon^2\right]=\mathbb{E}\left[\left(\varepsilon+\beta_1\left(X_1-\rho X_2-\left(1-\rho^2\right)^{1/2}\nu\right)\right)^2-\varepsilon^2\right]\\
    &=\beta_1^2\left(\mathrm{Var}\left[X_1-\rho X_2\right]+\left(1-\rho^2\right)\mathrm{Var}\left[\nu\right]\right)=2\beta_1^2\left(1-\rho^2\right).
\end{aligned}
\end{equation*}
Similarly, the VIMs for $X_2$ and $X_3$ are ${CP}_2=2\beta_2^2\left(1-\rho^2\right)$ and ${CP}_3=2\beta_3^2$, respectively.
\par
Since $f_{j,ALE}\left(x_j\right)=\beta_j x_j$ by the additive recovery property (\cite{Apley2020}), the ALE main effect VIM for $X_j$ is ${ALE}_{j,M}=\mathrm{Var}\left[f_{j, ALE}\left(X_j\right)\right]=\beta_j^2$. By Theorems 1 and 2, since there are no interactions, the ALE total effect VIMs are ${ALE}_{j,T}^Q={ALE}_{j,T}^S={ALE}_{j,M}=\beta_j^2$.
\end{appendices}

\end{document}